\documentclass[journal,twocolumn,10pt]{IEEEtran}
\usepackage{multirow}
\usepackage{textcomp}
\usepackage{stfloats}
\usepackage{url}
\usepackage{graphicx}
\usepackage{verbatim}
\usepackage{upgreek}
\usepackage{amssymb,amsmath}
\usepackage{color}
\usepackage{epstopdf}
\usepackage{bm}
\usepackage{cite}
\usepackage{array,color}
\usepackage{algorithm}
\usepackage{algpseudocode}
\usepackage{amsmath}
\usepackage{amsfonts}
\usepackage{amsthm}
\usepackage{graphics}
\usepackage{epsfig}
\usepackage{soul}
\usepackage{gensymb}
\soulregister\cite7
\soulregister\ref7
\usepackage{subfigure}
\usepackage{setspace}
\usepackage{makecell}
\newtheorem{thm}{Theorem}
\newtheorem{rmk}{Remark}

\newtheorem{lemma}{Lemma}
\begin{document}

\title{Doppler-Resilient OFDM Transmission for Low-Altitude Wireless Networks: Leveraging Affine-Frequency-Domain Pilot}

\author{Shuntian~Tang, Zesong~Fei,~\IEEEmembership{Senior~Member,~IEEE}, Xinyi~Wang,~\IEEEmembership{Member,~IEEE}, Le~Zhao, Yuanhan~Ni,~\IEEEmembership{Member,~IEEE}, Zilong~Liu,~\IEEEmembership{Senior~Member,~IEEE}

\thanks{Shuntian Tang, Zesong Fei, Xinyi Wang, and Le Zhao are with the School of Information and Electronics, Beijing Institute of Technology, Beijing 100081, China (E-mail:	tst491961069@163.com, feizesong@bit.edu.cn, wangxinyi@bit.edu.cn, tobin\_bit@icloud.com).}

\thanks{Yuanhan Ni is with the School of Electronic and Information Engineering, Beihang University, Beijing 100191, China (E-mail: yuanhanni@buaa.edu.cn).}

\thanks{Zilong Liu is with the School of Computer Science and Electronics Engineering, University of Essex, CO4 3SQ Colchester, U.K (E-mail: zilong.liu@essex.ac.uk).}

%\thanks{Fan Liu is with the National Mobile Communications Research Laboratory, Southeast University, Nanjing 210096, China (E-mail: fan.liu@seu.edu.cn).}

}

%\markboth{Submitted to IEEE Transactions on Vehicular Technology,~Vol.~xx, No.~xx, xxx~2022}%

%\IEEEpubid{0000--0000/00\$00.00~\copyright~2022 IEEE}
%\IEEEpubidadjcol

\maketitle

\begin{abstract}
Low-altitude wireless networks (LAWNs) built upon the conventional Orthogonal Frequency Division Multiplexing (OFDM) suffer from severe Doppler-induced performance degradation. While Affine Frequency Division Multiplexing (AFDM) is Doppler-resilient, it needs to cope with the relatively high channel estimation and data detection complexity as well as inflexibility in frequency-domain resource allocation. To address these issues, we introduce a novel affine frequency (AF) domain pilot embedding scheme to \textcolor{black}{enhance the robustness of OFDM systems} in high-mobility scenarios. To further improve the channel estimation accuracy, we investigate the mutual incoherence property (MIP) for pilot design and develop a Riemannian conjugate gradient (RCG)-based optimization scheme to minimize the mutual coherence of the dictionary atoms. \textcolor{black}{Furthermore, by revealing and exploiting the autoregressive (AR) properties of adjacent channel states, we develop a temporal convolutional network (TCN)-based predictor to replace conventional interpolation methods, thereby reducing the channel aging effect in high-mobility scenarios. Simulation results demonstrate that the proposed transmission scheme significantly improves channel estimation accuracy and consequently achieves superior bit error rate (BER) performance compared with conventional OFDM under high-mobility scenarios,} offering a promising paradigm for LAWN.
\end{abstract}

\begin{IEEEkeywords}
Affine frequency division multiplexing (AFDM), low-altitude wireless network (LAWN), Riemannian conjugate gradient (RCG), temporal convolutional network (TCN).
\end{IEEEkeywords}

\section{Introduction}

\IEEEPARstart{W}{ith} the rapid development of unmanned aerial vehicles (UAVs), low-altitude wireless networks (LAWNs) have attracted increasing attention in recent years~\cite{Luo2026WC}. Compared with conventional terrestrial networks, LAWNs can be flexibly deployed to provide seamless wireless connectivity, making them highly suitable for applications such as emergency rescue, disaster response, and aerial tourism~\cite{Qiu2019WC}. However, the high mobility of UAVs may lead to severe Doppler shifts and rapidly time-varying channels, which significantly degrade the communication reliability and quality of service (QoS)~\cite{Xu2026IOTM}.

To address this challenge, a number of advanced waveform schemes with enhanced Doppler resilience have been developed. Among many others, orthogonal time frequency space (OTFS) and affine frequency division multiplexing (AFDM) are regarded as two representative waveform  candidates for high-mobility communications due to their inherent robustness against Doppler. By representing wireless channels in the delay-Doppler (DD) or affine frequency (AF) domains, these waveform schemes have been demonstrated strong resilience to Doppler-induced channel variations, thereby achieving superior performance compared with conventional orthogonal frequency division multiplexing (OFDM) in rapidly time-varying channels~\cite{Sui2025arXiv}. 

Compared with the two-dimensional (2D) DD domain signal representation in OTFS, AFDM employs tunable chirp parameters to transform multiple orthogonal constant-frequency subcarriers into multiple orthogonal chirp waveforms, providing excellent backwards-compatibility with OFDM systems~\cite{Bemani2023TWC}. In~\cite{Bao2024PIMRC}, the authors investigated the impact of AFDM parameter configurations on communication performance, providing valuable insights into parameter selection and diversity enhancement. To further improve spectral efficiency and transmission reliability, AFDM-based index modulation schemes were investigated in~\cite{Liu2025TWC} and~\cite{Liu2026WCL}. Moreover, extensive studies have been conducted for multiple-input multiple-output AFDM (MIMO-AFDM) systems. Specifically, authors in \cite{Luo2025TWC} proposed a joint data detection and decoding receiver based on sparse graph theory and variational Bayesian inference, while~\cite{Sui2025TWC} developed a generalized spatial modulation aided AFDM (GSM-AFDM) framework, which demonstrated superior bit error rate (BER) performance compared with conventional AFDM systems.
	
Although AFDM has demonstrated great potential in various application scenarios, the relatively high channel estimation and data detection complexity still poses significant challenges for users with limited computational capabilities~\cite{Li2026IOT}. Furthermore, compared with OFDM, AFDM may not allow highly flexible scheduling across orthogonal resource elements (REs)~\cite{Gao2026TCom}. Due to the matured digital signal processing  implementation and with the recent consensus in the third generation partnership project (3GPP) meetings, OFDM is therefore expected to remain as the dominant waveform for the next-generation wireless systems.
Motivated by these considerations, a promising approach is not to completely replace OFDM with AFDM, but rather to incorporate partial AF-domain processing into OFDM systems to enhance their robustness against Doppler effects. \textcolor{black}{Unlike conventional OFDM channel estimation, which only estimates the diagonal elements of the frequency-domain channel matrix and neglects Doppler-induced inter-carrier interference (ICI), AF-domain channel estimation explicitly estimates the delay and Doppler parameters of each propagation path and reconstructs the channel state information (CSI)~\cite{Bemani2023TWC}, enabling an explicit characterization of Doppler effects. Motivated by this observation, introducing AF-domain channel estimation into OFDM systems enables the corresponding systems to effectively handle Doppler effects~\cite{tang2026dopplerresilientleosatelliteofdm}}. In such a way, one can preserve the key advantages of conventional OFDM systems, including flexible time-frequency resource allocation and low-complexity frequency-domain equalization. Meanwhile, it overcomes a fundamental limitation of conventional frequency-domain channel estimation, where Doppler-induced ICI cannot be accurately characterized. By incorporating AF-domain estimation, Doppler effects can be more effectively captured, thereby significantly improving OFDM performance in high-mobility environments. Despite certain loss of multipath diversity in uncoded systems~\cite{Bemani2023TWC}, such a trade-off is useful for balancing Doppler robustness, implementation complexity, and compatibility with existing OFDM systems.

As an important advantage of AF-domain channel estimation, the effective channel exhibits inherent sparsity, enabling the application of compressive sensing (CS)-based estimation algorithms~\cite{Cao2024ICCT,Wu2025ICC}. However, in practical high-mobility environments, fractional Doppler shifts introduce strong correlations among dictionary atoms, which significantly degrades sparse recovery performance and may even lead to estimation failure.
To address this issue, one possible approach is to improve the channel estimation algorithm at the cost of increased computational complexity. For example, \cite{Mamandipoor2016TSP} and~\cite{Li2021JSAC} proposed enhanced orthogonal matching pursuit (OMP)-based frameworks, where Newton-based refinement techniques were incorporated to compensate for the discrete estimation grid mismatch. In~\cite{Wei2022TWC}, the authors proposed an off-grid sparse Bayesian learning (SBL)-based estimator, in which the off-grid components were treated as hyperparameters and iteratively refined. Another approach is to optimize the pilot design to reduce the mutual coherence of the dictionary while maintaining low algorithmic complexity~\cite{Wang2021WCL,Mohammadian2017CL}. Considering the complexity constraints of practical deployment, as well as the limited existing research on AF-domain pilot design, this paper adopts the latter approach.

Moreover, independent pilot symbols are employed in the proposed framework, where pilot and data symbols are transmitted in different OFDM symbols. As a result, under severe Doppler conditions, directly using the estimated CSI for subsequent data equalization may suffer from serious CSI aging effects, even when interpolation techniques are applied~\cite{Wu2021JSAC}. To mitigate this issue, neural network (NN)-based CSI prediction methods were investigated in~\cite{Li2020WCL,Zhang2026TWC} to predict future CSIs in various systems. 
\textcolor{black}{As a lightweight alternative, pilot prediction techniques exploit the received pilots from a subset of symbols to reconstruct the pilot observations that would have been obtained if pilot symbols had been transmitted across the entire slot. The reconstructed pilot observations are then used to estimate the CSI of all symbols within the slot~\cite{tang2026dopplerresilientleosatelliteofdm}. By shifting the prediction objective from the full CSI to pilot observations, the resulting models become significantly more lightweight in AFDM systems as the pilot dimension is much smaller than that of the entire CSI in the AF domain. However, the LSTM-based predictor in~\cite{tang2026dopplerresilientleosatelliteofdm} restricted the pilot symbols to occupy the initial position of each slot. This constraint eliminates pilot deployment flexibility and hinders compatibility with dynamic resource scheduling in practical systems.}

In this paper, we propose an AF-domain pilot assisted OFDM transmission framework for high-mobility LAWNs to enhance the Doppler robustness of conventional OFDM systems. We first establish the relationship between CSI in the frequency and AF domains, based on which an AF-domain pilot structure and the corresponding Doppler-resilient OFDM system model are developed to integrate AFDM-related techniques into OFDM systems.
To improve channel estimation performance, we further develop a transmit pilot optimization scheme based on the Riemannian conjugate gradient (RCG) algorithm. The proposed scheme minimizes the mutual coherence of the channel estimation dictionary, thereby enhancing sparse recovery capability and channel estimation accuracy.
\textcolor{black}{Furthermore, to mitigate channel aging and enable flexible pilot placement}, we investigate the autoregressive (AR) characteristics of AF-domain channel evolution and develop a temporal convolutional network (TCN)-based pilot prediction scheme. 
%Extensive simulation results demonstrate that the proposed framework significantly improves BER and channel estimation performance compared with conventional OFDM pilot schemes in severe Doppler scenarios.

The main contributions are summarized as follows:
\begin{itemize}
	\item[$\bullet$] We analyze the CSI representations in both the frequency and AF domains, and establish the mathematical foundation for reconstructing the frequency-domain CSI from AF-domain channel estimation results. Based on this analysis, we propose a novel AF-domain pilot-assisted OFDM transmission framework for high-mobility LAWNs. Unlike conventional OFDM pilots that only estimate the diagonal elements of the frequency-domain channel matrix, the proposed AF-domain pilot can effectively characterize Doppler-induced ICI on a path-wise basis through AF-domain estimation, thereby achieving enhanced robustness under severe Doppler conditions.
	
	\item[$\bullet$] We develop an RCG-based AF-domain pilot optimization scheme by minimizing the mutual coherence of the \textcolor{black}{channel estimation} dictionary. Specifically, the pilot design problem is formulated as a manifold optimization problem under constant-modulus constraints, and the RCG algorithm is employed to iteratively optimize the pilot sequence. Owing to the optimized AF-domain pilot design, the proposed scheme can effectively reduce dictionary coherence and alleviate off-grid effects, significantly enhancing sparse channel estimation performance without increasing the receiver complexity.
	
	\item[$\bullet$] We prove the inherent AR structure of the received pilot symbol sequence and develop a TCN-based NN architecture for pilot prediction. By leveraging the temporal evolution characteristics of pilot symbols, the proposed schemes can \textcolor{black}{support flexible pilot placement, and} effectively compensate for channel variation caused by high Doppler shifts, thereby reducing channel estimation errors in dynamic environments compared with conventional interpolation \textcolor{black}{and LSTM-based counterpart in~\cite{tang2026dopplerresilientleosatelliteofdm} under the same pilot overhead.}
	%In addition, the computational complexity of the proposed prediction networks is analyzed to evaluate their practical feasibility.
\end{itemize}

The remainder of this paper is organized as follows. Section~\ref{Preliminaries} presents the proposed AF-domain pilot-assisted OFDM framework. Section~\ref{Transmit pilot design} develops the channel estimation dictionary design criterion and the corresponding RCG-based pilot optimization scheme. Section~\ref{Receive sequence prediction} investigates the AR characteristics of wireless channels and proposes a TCN-based predictor to mitigate channel aging. Section~\ref{Simulation Results} presents simulation results, and Section~\ref{Conclusion} concludes the paper.

%The remainder of this paper is organized as follows. Section~\ref{Preliminaries} first analyzes the impact of high Doppler shifts on OFDM systems, introduces the proposed AF-domain pilot, and presents the corresponding OFDM-enhanced system model. Section~\ref{Transmit pilot design} introduces the channel estimation dictionary design criterion and develops an RCG-based pilot optimization scheme. In Section~\ref{Receive sequence prediction}, the AR characteristics of wireless channels are proved, based on which TCN-based pilot prediction NN is developed to mitigate the channel aging effect. Section~\ref{Simulation Results} provides simulation results to validate the effectiveness of the proposed transmission framework. Finally, Section~\ref{Conclusion} concludes the paper.

\emph{Notations}: Unless otherwise specified, we use a lowercase letter $a$, a boldface lowercase letter $\mathbf{a}$, a boldface capital letter $\mathbf{A}$, and a calligraphy letter $\mathcal{A}$ to denote a scalar, a vector, a matrix, and a set, respectively. $\text{diag}\left(\cdot\right)$ denotes the diagonal operation; $\left(\cdot\right)^*$, $\left(\cdot\right)^T$, and $\left(\cdot\right)^H$ denote the conjugate,  transpose, and Hermitian transpose operations, respectively; $\mathbb{R}$ and $\mathbb{C}$ denote the real and complex number fields, respectively; $\operatorname{Re}\{\cdot\}$ and $\operatorname{Im}\{\cdot\}$ denote real and imaginary parts of its argument, respectively; $\mathbf{1}_{N}$ and $\mathbf{I}_{N}$ denote the all-one vector of size $N \times 1$ and identity matrix of size $N \times N$, respectively.
% $\left(\cdot\right)_N$ denotes the modulo $N$ operation; 

\section{AF-Domain Pilot Aided OFDM Transmission Framework}
\label{Preliminaries}
In this section, we first analyze the impact of high Doppler on conventional OFDM systems. Then, to address the resulting performance degradation in LAWN scenarios, we propose an AF-domain pilot assisting scheme that enhances the Doppler resilience of OFDM.
\subsection{Analysis of the Impact of High Doppler on OFDM}
We begin by introducing the low-altitude UAV–ground communication channel model. Due to the high mobility of UAVs, the corresponding channel is generally time–frequency doubly selective, and the channel impulse response (CIR) can be expressed as~\cite{Bemani2023TWC}
\begin{equation} \label{CIR}
	h(t,\tau)=\sum_{i=0}^{P-1}h_ie^{-j2\pi f_it}\delta(\tau-\tau_i),
\end{equation}
where $P$ denotes the number of propagation paths, and $h_i$, $l_i$ and $f_i$ represent the complex channel gain, propagation delay, and Doppler shift of the $i$-th path, respectively.

Consider an OFDM system with $N$ subcarriers, subcarrier spacing $\Delta f$, and $L$ OFDM symbols per slot. Under the assumption that the channel delay and Doppler parameters remain approximately constant within each OFDM slot, the frequency-domain channel matrix corresponding to the $k$-th $(k=0,\dots,L-1)$ OFDM symbol can be derived from the CIR in (\ref{CIR}) as~\cite{Ma2025WCL}
\begin{equation} \label{CFR}
	\mathbf{H}_{\text{FD},k}=\sum_{i=0}^{P-1}\tilde{h}_{i,k}\mathbf{F}\mathbf{\Delta}_{f_i}\mathbf{\Pi}^{l_i}\mathbf{F}^H,
\end{equation}
where $\mathbf{F}$ denotes the normalized $N$-point DFT matrix, $\mathbf{\Delta}_{f_i} = \text{diag}\left([1,e^{-j2\pi f_i},\cdots,e^{-j2\pi f_i\left(N-1\right)}]\right)$ is the Doppler shift diagonal matrix, $\mathbf{\Pi}^{l_i}$ denotes the cyclic permutation matrix associated with the normalized delay $l_i=\tau_i f_s=\tau_i\cdot N\Delta f$, and $\tilde{h}_{i,k}$ denotes the time-varying complex channel gain which is given by 
\begin{equation} \label{htilde}
	\tilde{h}_{i,k}={h}_{i}e^{-j2\pi f_c/f_s l_i}e^{-j2\pi \nu_i[(k+1)L_\text{CP}+kN]/N},
\end{equation}
where $f_c$ denotes the radio frequency (RF) carrier frequency, $\nu_i = N f_i$ denotes the normalized Doppler shift, and $L_\text{CP}$ denotes the length of cyclic prefix (CP).

\begin{figure*}[t]
	\centering
	\subfigure[No Doppler ($\nu_{\max}=0$).]{
		\label{1}
		\includegraphics[width=0.3\textwidth]{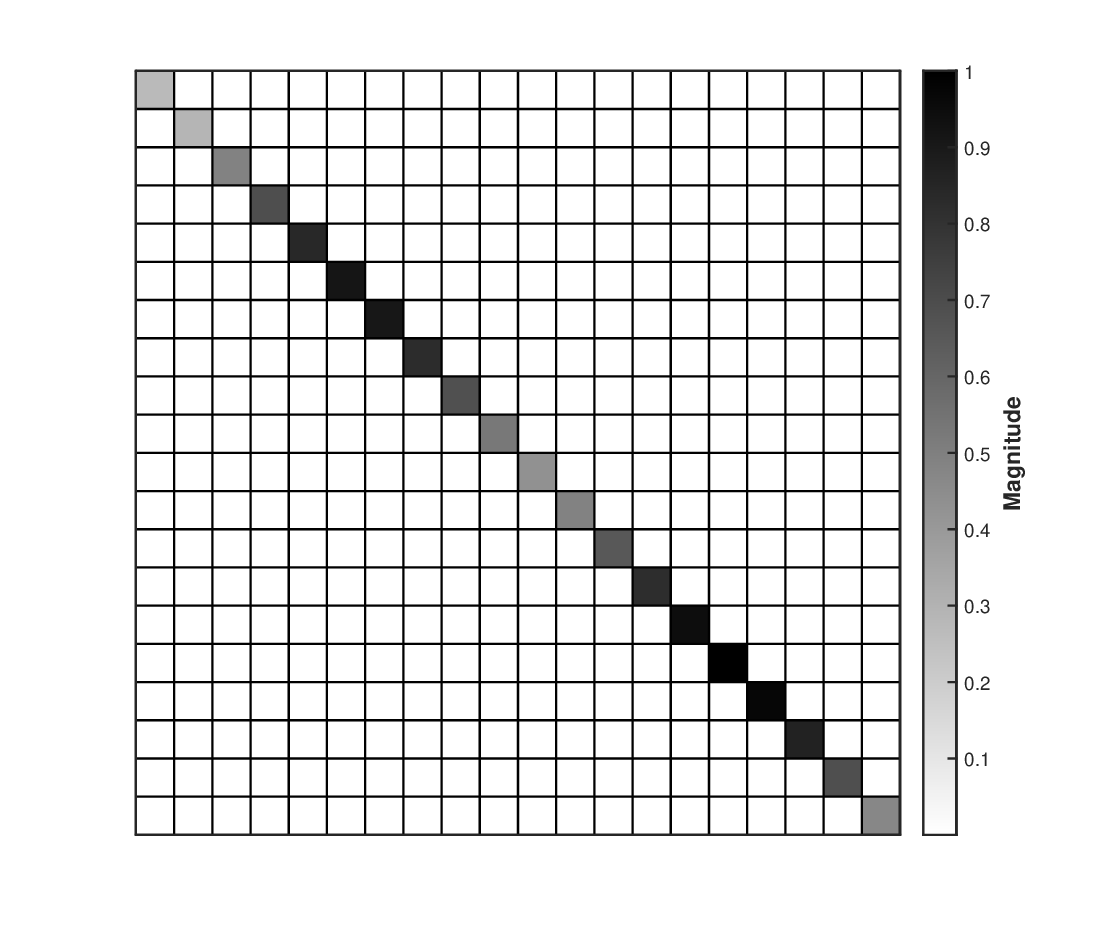}}
	\quad
	\subfigure[Integer Doppler only ($\nu_{\max}=2$).]{
		\label{2}
		\includegraphics[width=0.3\textwidth]{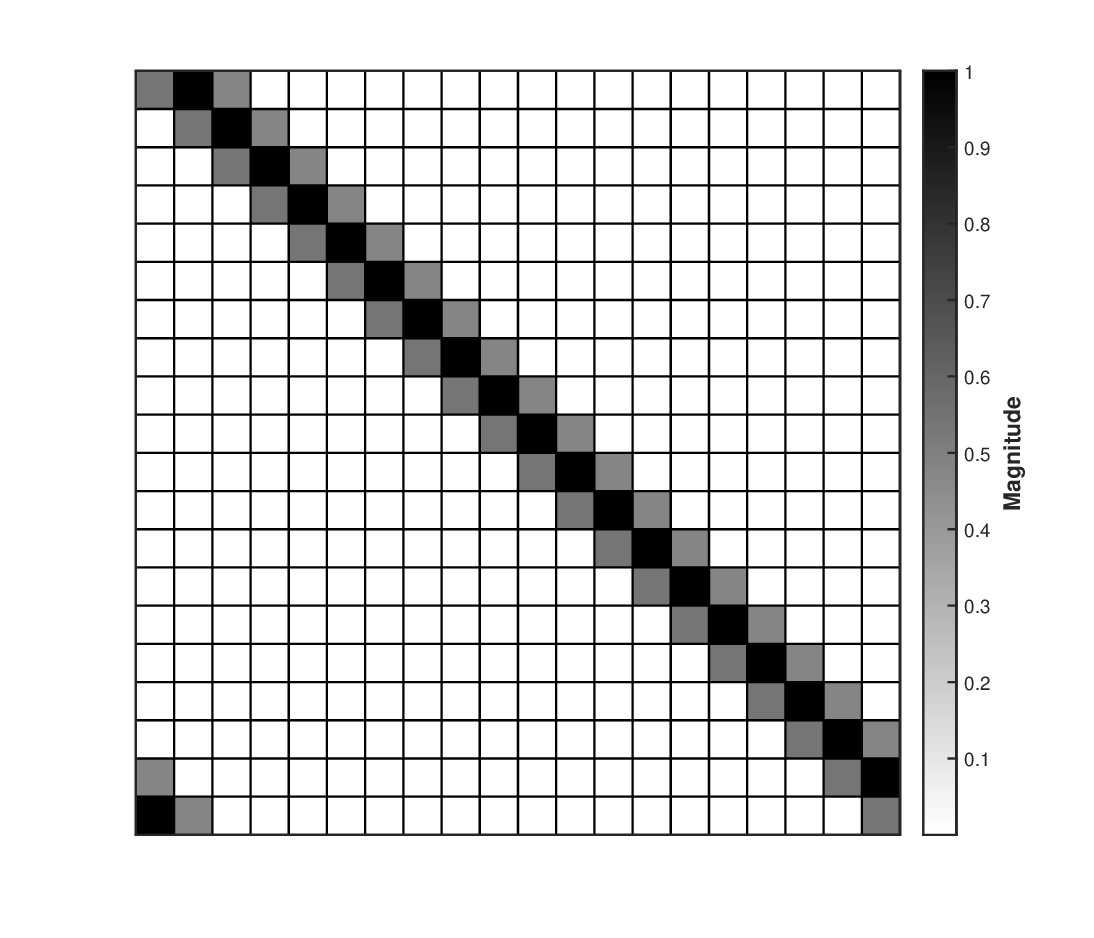}}
		\quad
	\subfigure[Fractional Doppler present ($\nu_{\max}=0.5$).]{
		\label{3}
		\includegraphics[width=0.3\textwidth]{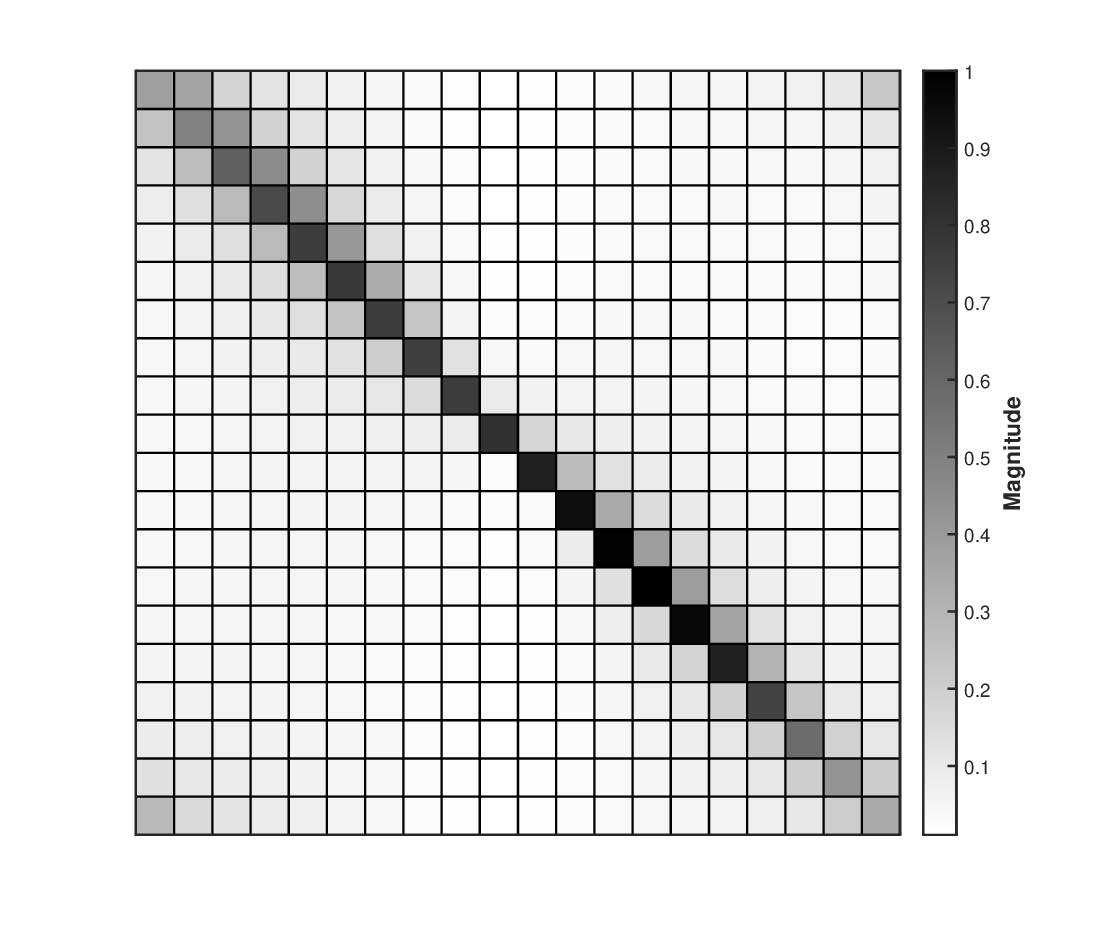}}
	\caption{Frequency-domain channel matrix representation under various Doppler conditions.}
	\label{fig:Doppler impact}
\end{figure*}

%Let $\nu_i = \alpha_i + a_i$ characterize the Doppler effect, where $\alpha_i \in [-\alpha_{\max}, \alpha_{\max}]$ is the integer component while $a_i \in (-0.5, 0.5]$ is the fractional component. Under high-mobility conditions, fractional Doppler $a_i$ destroys subcarrier orthogonality and spreads energy across adjacent subcarriers, making single-tone pilots unable to capture the true channel response. Meanwhile, the channel varies rapidly from symbol to symbol; hence, interpolation between pilot symbols becomes unreliable. As a result, conventional OFDM channel estimation suffers severe degradation in the LEO satellite system.
Let $\nu_i = \alpha_i + a_i$ characterize the Doppler effect, where $\alpha_i \in [-\alpha_{\max}, \alpha_{\max}]$ is the integer component that induces cyclic shifts across subcarriers without destroying their orthogonality, and $a_i \in (-0.5, 0.5]$ is the fractional component that causes energy leakage into other subcarriers, resulting in inter-carrier interference (ICI), as illustrated in Fig.~\ref{fig:Doppler impact}. Conventional OFDM systems operate under the assumption of (near-)orthogonal subcarriers, which leads to a predominantly diagonal frequency-domain channel matrix $\mathbf{H}_{\text{FD},k}$. Under this assumption, full CSI can be readily acquired using pilots placed on individual subcarriers. However, in the presence of fractional Doppler shifts, subcarrier orthogonality is destroyed and the channel energy spreads beyond the main diagonal. Consequently, pilots on individual subcarriers are no longer sufficient to characterize the true channel response, resulting in inevitable degradation in channel estimation and equalization performance.

\begin{rmk}
	According to~\cite{3gpp.38.104}, the carrier frequency accuracy requirements in NR systems ensure that large residual carrier frequency offsets (CFOs) can be effectively mitigated through synchronization. As a result, the dominant integer Doppler components can be largely compensated after synchronization. Therefore, in practical systems, channel estimation and equalization mainly need to focus on the \textbf{fractional Doppler} components, which remain the primary source of ICI.
\end{rmk}

In UAV–ground communications, the time variation of $\tilde{h}_{i,k}$ in (\ref{htilde}) causes frequency-domain channel matrices to vary across OFDM symbols, even when the delay and Doppler parameters remain approximately constant. Conventional OFDM channel estimation typically relies on pilot interpolation to obtain CSI on data symbols, which is effective in low-mobility or single-path scenarios but becomes insufficient in high-mobility multipath environments. In such cases, the channel evolution is governed by multiple rapidly varying exponential components, which cannot be accurately captured by interpolation, leading to severe estimation errors.

%In UAV–ground communications, due to the time-varying nature of $\tilde{h}_{i,k} $ in (\ref{htilde}), the frequency-domain channel matrices vary from one OFDM symbol to another, even within the condition that the channel delay and Doppler parameters remain approximately constant. Conventional OFDM channel estimation relies on interpolating pilot-based estimates to obtain the CSI on data symbols. While this approach is effective in low-mobility or single-path scenarios, it becomes inadequate in high-mobility multipath environments. In such cases, the channel evolution is governed by a superposition of multiple rapidly varying exponential components, which cannot be accurately captured by interpolation, leading to significant estimation errors.

%It can be seen that the degradation of OFDM channel estimation in LEO satellite systems stems from two challenges: 
%\begin{itemize}
%	\item[$\bullet$] Fractional Doppler destroys subcarrier orthogonality, causing energy to spread off the diagonal so that pilot tones on individual subcarriers can no longer capture the overall channel information.
%	\item[$\bullet$] Rapid symbol-to-symbol channel variation under high-mobility multipath renders interpolation-based CSI approximation unreliable.
%\end{itemize}
\subsection{Doppler-Resilient OFDM Transmission with AF-Domain Pilot}
To address the challenges caused by fractional Doppler shifts and interpolation errors in UAV–ground communications, we introduce an AF-domain-pilot-assisted transmission scheme to enhance Doppler resilience.

We first briefly review the fundamental principles of the AF-domain pilot~\cite{tang2026dopplerresilientleosatelliteofdm}. 
The input-output relationship of the $k$-th symbol per slot in the AF domain can be expressed in matrix form as
\begin{equation} \label{IO-relation}
	\mathbf{y}_k =\sum_{i=1}^{P}\tilde{h}_{i,k}\mathbf{H}_i\mathbf{x}+\widetilde{\mathbf{w}}= \mathbf{H}_{\text{AFD},k}\mathbf{x}+\widetilde{\mathbf{w}},
\end{equation}
with 
\begin{equation}
	\mathbf{H}_i=\mathbf{\Lambda}_{c_2}\mathbf{F}\mathbf{\Lambda}_{c_1}\mathbf{\Gamma}_i\mathbf{\Delta}_{f_i}\mathbf{\Pi}^{l_i}\mathbf{\Lambda}_{c_1}^H\mathbf{F}^H\mathbf{\Lambda}_{c_2}^H \in \mathbb{C}^{N \times N},
\end{equation}
where $c_1$ and $c_2$ are tunable parameters designed to achieve full diversity over doubly selective fading channels, $\mathbf{\Lambda}_{c}=\text{diag}\left([1,e^{-j2\pi c},\cdots,e^{-j2\pi c\left(N-1\right)^2}]\right)$, $\widetilde{\mathbf{w}} \sim \mathcal{CN}(0, \sigma^2 \mathbf{I}_N)$ denotes the additive white Gaussian noise (AWGN), and $\mathbf{\Gamma}_i$ is an $N$-order diagonal matrix given by 
\begin{equation}
	\mathbf{\Gamma}_{i}(i,i)= \begin{cases}e^{-j 2 \pi c_1\left(N^2-2 N\left(l_i-n\right)\right)}, & n<l_i, \\ 1, & n \geq l_i.\end{cases}
\end{equation}
Note that when $2Nc_1$ is an integer value and $N$ is even, $\mathbf{\Gamma}_i=\mathbf{I}_N$.

% In practical applications, $N$ is typically chosen to be even to facilitate the fast Fourier transform (FFT) implementation; hence, $2Nc_1$ is usually an integer based on (\ref{c1}). As a result, $\mathbf{\Gamma}_i$ simplifies to $\mathbf{I}_N$, and the the AF-domain channel matrix becomes
In practical applications, subcarrier number $N$ is typically chosen to be even to facilitate the fast Fourier transform (FFT) implementation, and $2Nc_1$ is usually set as an integer to achieve full diversity gain~\cite{Bemani2021ICC}.
As a result, $\mathbf{\Gamma}_i$ simplifies to $\mathbf{I}_N$, and the AF-domain channel matrix becomes
\begin{equation} \label{CAFR}
	\mathbf{H}_{\text{AFD},k}=\sum_{i=1}^{P}\tilde{h}_{i,k}\mathbf{\Lambda}_{c_2}\mathbf{F}\mathbf{\Lambda}_{c_1}\mathbf{\Delta}_{f_i}\mathbf{\Pi}^{l_i}\mathbf{\Lambda}_{c_1}^H\mathbf{F}^H\mathbf{\Lambda}_{c_2}^H.
\end{equation}

By defining the unitary transform matrix $\mathbf{T}=\mathbf{\Lambda}_{c_2}\mathbf{F}\mathbf{\Lambda}_{c_1}\mathbf{F}^H$, the relationship between the frequency-domain channel matrix (\ref{CFR}) and AF-domain channel matrix (\ref{CAFR}) is given by
\begin{equation} \label{AO-relation}
	\mathbf{H}_{\text{FD},k}=\mathbf{T}^H\mathbf{H}_{\text{AFD},k}\mathbf{T}.
\end{equation}
This relationship indicates that the frequency-domain CSI can be directly obtained from the AF-domain CSI through a unitary similarity transformation. Unlike conventional OFDM channel estimation schemes, AF-domain channel estimation explicitly estimates the path gains $\tilde{h}_{i,k}$, delays $l_i$, and Doppler shifts $\nu_i$ for each path and then reconstructs the full channel matrix~\cite{Bemani2023TWC}. This sensing-like procedure provides a more complete and structurally faithful representation of the doubly selective channel. 

\begin{figure*}[t]
	\centering
	\includegraphics[width=0.85\textwidth]{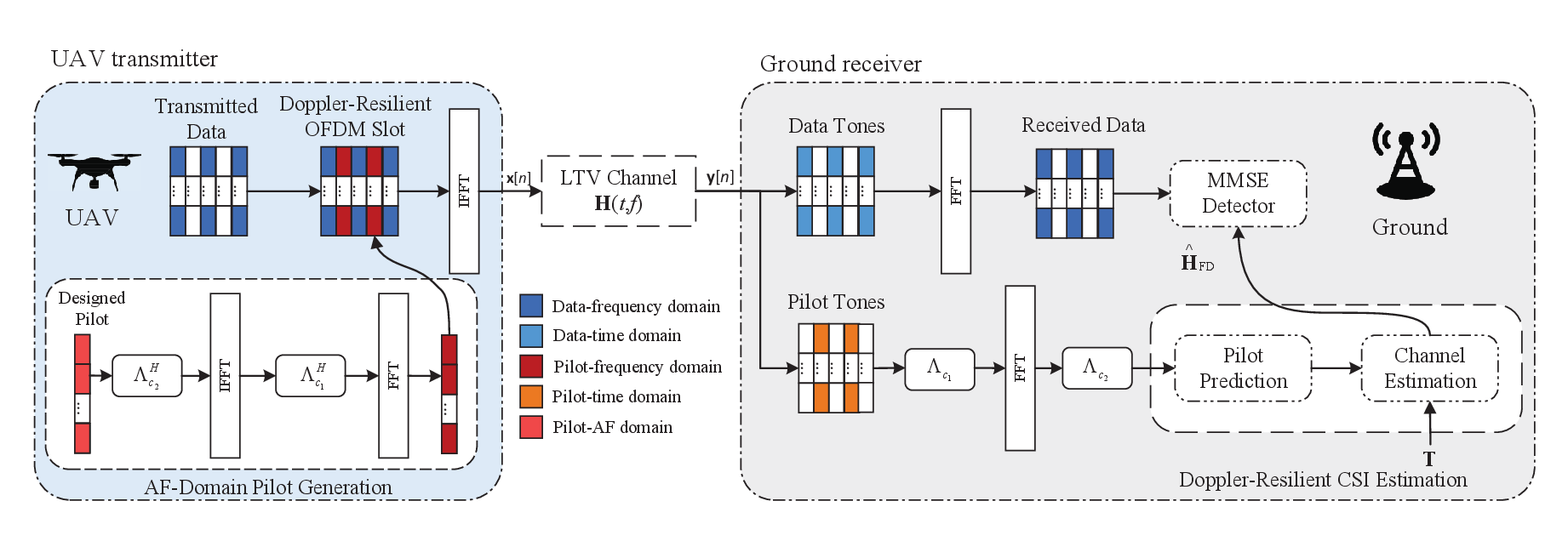}
	\caption{Block diagram of AF-domain pilot embedding enhanced OFDM system.}
	\label{fig:sys}
\end{figure*}
Motivated by this observation, we propose an AF-domain channel estimation enhanced Doppler-resilient OFDM transmission framework, as illustrated in Fig.~\ref{fig:sys} at the top of the next page. In the proposed framework, the conventional OFDM pilots are replaced by AF-domain pilots, which are constructed by sequentially applying the IDAFT and FFT to the pilot sequence designed in Section~\ref{Transmit pilot design}. These pilots are embedded at predetermined positions within each slot, providing the basis for the pilot prediction process described in Section~\ref{Receive sequence prediction} at the receiver. 
%The corresponding slot structure will be specified in Section~\ref{Receive sequence prediction}.
The AF-domain pilot sequence exhibits an AR structure, as discussed in Section~\ref{Receive sequence prediction}, which enables the inference of remaining received pilots from only a subset of the observed pilots. This implies that, for resource elements originally occupied by data symbols, the corresponding received pilot symbols can be inferred as if pilot symbols had been transmitted at those positions, thereby enabling the estimation of the associated CSI. Based on these inferred \textbf{virtual pilots}, channel estimation is then performed in the AF domain, and the interpolation process is eliminated, thus avoiding interpolation errors. Subsequently, the frequency-domain channel matrix is reconstructed using the relation in (\ref{AO-relation}). In what follows, we detail the designs of the transmit pilot sequence and the receiver-side pilot prediction modules, respectively.

\begin{rmk}
	Although the reconstructed frequency-domain channel matrix is two-dimensional according to (\ref{AO-relation}), it remains \textbf{diagonally dominant} after the integer Doppler component is eliminated through frequency synchronization. Such matrices can be efficiently inverted using low-complexity iterative algorithms, such as the Jacobi or Gauss–Seidel methods, resulting in lower computational cost compared with equalization performed directly in the AF domain.
\end{rmk}

\section{AF-Domain Pilot Optimization based on Riemannian Manifold}
\label{Transmit pilot design}
In this section, we develop an AF-domain pilot sequence optimization framework to improve the channel estimation accuracy. First, a pilot sequence optimization problem is established to minimize the mutual incoherence property (MIP) of channel estimation dictionary. Since this problem admits no closed-form solution~\cite{Wang2021WCL}, the RCG method is employed to obtain an efficient numerical solution.
%\subsection{AF-Domain Pilot Sequence Optimization Problem}
%\subsection{Sparse Channel Estimation Model for AFDM}

%According to \cite{Bemani2023TWC}, the AF-domain channel matrix $\mathbf{H}_{\text{eff}}$ exhibits inherent sparsity when the normalized Doppler shifts of all propagation paths are integers. When the normalized Doppler shifts are fractional, the magnitude $\vert \mathbf{H}_i(p, q) \vert$ decays rapidly as $q$ deviates from $\left(p + \operatorname{loc}_i\right)_N$. Consequently, the channel matrix can still be regarded approximately sparse.
To exploit this sparsity in the AF domain for channel estimation, we reformulate (\ref{IO-relation}) with respect to the pilot symbols $\mathbf{x}_\text{p}$ in each slot as
\begin{equation} \label{liner_IO}
	\mathbf{y}_k= \sum_{i=1}^{P}\tilde{h}_{i,k}\mathbf{H}_i\mathbf{x}_\text{p}+\widetilde{\mathbf{w}} \triangleq \mathbf{\Phi}\mathbf{h} + \widetilde{\mathbf{w}},
\end{equation}
where $\mathbf{\Phi}=[\mathbf{H}_1(l_1,v_1)\mathbf{x}_\text{p}|\cdots|\mathbf{H}_M(l_M,v_M)\mathbf{x}_\text{p}] \in \mathbb{C}^{N \times M}$ denotes the channel dictionary matrix that contains all possible delay–Doppler tap combinations $(l_i, v_i)$, resulting in a total of $M$ atoms, and the vector $\mathbf{h} \in \mathbb{C}^{M \times 1}$ is the corresponding sparse fading coefficient vector. 
Notably, only $P$ elements in $\mathbf{h}$ are non-zero and the channel estimation problem in (\ref{liner_IO}) thus can be naturally formulated as a sparse signal recovery problem. Both conventional CS techniques~\cite{Zhang2021ICSPCC} and SBL-based methods~\cite{Tang2025ICCC} can be employed to efficiently solve this problem.
Notably, the performance of sparse recovery algorithms critically depends on the channel dictionary matrix $\mathbf{\Phi}$. High mutual coherence among its columns can significantly degrade sparse channel estimation accuracy~\cite{Mohammadian2017CL}. Since $\mathbf{\Phi}$ is determined by the AF-domain pilot sequence $\mathbf{x}_{\text{p}}$, we optimize the pilot design by minimizing the MIP of $\mathbf{\Phi}$, thereby enhancing channel estimation performance.
%Although sparse recovery algorithms can effectively exploit the inherent sparsity of the AF-domain channel, their estimation performance critically depends on the structure of the channel estimation dictionary matrix $\mathbf{\Phi}$. In particular, strong mutual coherence among the columns of $\mathbf{\Phi}$ may significantly degrade sparse recovery accuracy and convergence behavior~\cite{Mohammadian2017CL}. Since the dictionary matrix is directly determined by the AF-domain pilot sequence $\mathbf{x}_{\text{p}}$, an appropriate pilot design is essential for improving the robustness and accuracy of sparse channel estimation. In the following, we focus on optimizing the pilot sequence by minimizing the MIP of the channel dictionary matrix, thereby enhancing the performance of sparse channel estimation algorithms.
%\begin{rmk}
%	Increasing the sampling intervals $l_s$ and $\alpha_s$ can reduce the mutual coherence of the dictionary matrix $\mathbf{\Phi}$; however, overly coarse discretization may cause the true delay–Doppler components to fall off the predefined grid, leading to an off-grid mismatch and degraded estimation performance~\cite{Tang2025ICCC}. Therefore, a trade-off exists between reducing dictionary coherence and preserving accurate channel representation.
%\end{rmk}

Specifically, the MIP of the dictionary matrix $\mathbf{\Phi}$, denoted by $\rho(\mathbf{\Phi})$, is defined as the maximum absolute value of the normalized inner product between any two distinct columns of $\mathbf{\Phi}$~\cite{Chen2020ICC}, i.e.,
\begin{equation} \label{MIP1}
	\rho(\mathbf{\Phi})= \max_{1\leq i,j \leq M,m \neq n}\frac{\Vert\mathbf{\Phi}(:,i)^H\mathbf{\Phi}(:,j)\Vert_2}{\Vert\mathbf{\Phi}(:,i)\Vert_2\Vert\mathbf{\Phi}(:,j)\Vert_2},
\end{equation}
where $\mathbf{\Phi}(:,i)=\mathbf{H}_i\mathbf{x}_{\text{p}}$ denotes the $i$-th column of $\mathbf{\Phi}$. The MIP characterizes the correlation among the dictionary atoms, and a higher coherence generally leads to degraded sparse recovery accuracy and convergence performance. Furthermore, by substituting $\mathbf{\Phi}(:,i)=\mathbf{H}_i\mathbf{x}_{\text{p}}$ into (\ref{MIP1}), the MIP can be equivalently expressed as
%\begin{align} \label{MIP2}
%	\rho(\mathbf{\Phi})&= \max_{1\leq i,j \leq M,i \neq j}\sqrt{\frac{\mathbf{x}_{\text{p}}^H\mathbf{H}_j^H\mathbf{H}_i\mathbf{x}_{\text{p}}\mathbf{x}_{\text{p}}^H\mathbf{H}_i^H\mathbf{H}_j\mathbf{x}_{\text{p}}}{\mathbf{x}_{\text{p}}^H\mathbf{H}_i^H\mathbf{H}_i\mathbf{x}_{\text{p}}\mathbf{x}_{\text{p}}^H\mathbf{H}_j^H\mathbf{H}_j\mathbf{x}_{\text{p}}}} \\ \notag
%	&\triangleq \max_{1\leq i,j \leq M,i \neq j}\frac{\mathbf{x}_{\text{p}}^H\mathbf{H}_j^H\mathbf{H}_i\mathbf{x}_{\text{p}}\mathbf{x}_{\text{p}}^H\mathbf{H}_i^H\mathbf{H}_j\mathbf{x}_{\text{p}}}{\mathbf{x}_{\text{p}}^H\mathbf{H}_i^H\mathbf{H}_i\mathbf{x}_{\text{p}}\mathbf{x}_{\text{p}}^H\mathbf{H}_j^H\mathbf{H}_j\mathbf{x}_{\text{p}}} \\ \notag
%	&=\max_{1\leq i,j \leq M,i \neq j}\frac{\mathbf{x}_{\text{p}}^H\mathbf{H}_j^H\mathbf{H}_i\mathbf{x}_{\text{p}}\mathbf{x}_{\text{p}}^H\mathbf{H}_i^H\mathbf{H}_j\mathbf{x}_{\text{p}}}{\mathbf{x}_{\text{p}}^H\mathbf{x}_{\text{p}}\mathbf{x}_{\text{p}}^H\mathbf{x}_{\text{p}}}=\psi(\mathbf{x}_{\text{p}}),
%\end{align}
\begin{align} \label{MIP2}
	\rho(\mathbf{\Phi})&= \max_{1\leq i,j \leq M,i \neq j}\sqrt{\frac{\mathbf{x}_{\text{p}}^H\mathbf{H}_j^H\mathbf{H}_i\mathbf{x}_{\text{p}}\mathbf{x}_{\text{p}}^H\mathbf{H}_i^H\mathbf{H}_j\mathbf{x}_{\text{p}}}{\mathbf{x}_{\text{p}}^H\mathbf{H}_i^H\mathbf{H}_i\mathbf{x}_{\text{p}}\mathbf{x}_{\text{p}}^H\mathbf{H}_j^H\mathbf{H}_j\mathbf{x}_{\text{p}}}} \\ \notag
	&\triangleq\max_{1\leq i,j \leq M,i \neq j}\frac{\mathbf{x}_{\text{p}}^H\mathbf{H}_j^H\mathbf{H}_i\mathbf{x}_{\text{p}}\mathbf{x}_{\text{p}}^H\mathbf{H}_i^H\mathbf{H}_j\mathbf{x}_{\text{p}}}{\mathbf{x}_{\text{p}}^H\mathbf{x}_{\text{p}}\mathbf{x}_{\text{p}}^H\mathbf{x}_{\text{p}}}=\psi(\mathbf{x}_{\text{p}}),
\end{align}
which explicitly reveals that the MIP of $\mathbf{\Phi}$ is a function of the AF-domain pilot sequence $\mathbf{x}_{\text{p}}$.

%From (\ref{MIP2}), it is observed that the MIP of the dictionary matrix can be improved either by reducing the number of dictionary atoms or by appropriately designing the pilot sequence. However, decreasing the atom number generally requires coarser delay–Doppler discretization, which may introduce off-grid mismatch and degrade channel representation accuracy~\cite{Tang2025ICCC}. Therefore, in this work, we focus on pilot sequence optimization as a more effective and robust approach for improving the MIP of $\mathbf{\Phi}$.
From (\ref{MIP2}), the mutual incoherence property (MIP) of the dictionary matrix can be improved either by reducing the number of dictionary atoms or by designing the pilot sequence. However, reducing the atom number requires coarser delay–Doppler discretization, which may cause off-grid mismatch and degrade channel representation accuracy~\cite{Tang2025ICCC}. Therefore, this work focuses on pilot sequence design as a more effective and robust way to improve the MIP of $\mathbf{\Phi}$.
Naturally, based on (\ref{MIP2}), the AF-domain pilot sequence optimization problem can be formulated as
\begin{align} \label{P1}
	(\mathcal{P}1)\quad\min_{\mathbf{x}_{\text{p}}} \quad  &\max_{1\leq i,j \leq M,i \neq j}\frac{\mathbf{x}_{\text{p}}^H\mathbf{H}_j^H\mathbf{H}_i\mathbf{x}_{\text{p}}\mathbf{x}_{\text{p}}^H\mathbf{H}_i^H\mathbf{H}_j\mathbf{x}_{\text{p}}}{\mathbf{x}_{\text{p}}^H\mathbf{x}_{\text{p}}\mathbf{x}_{\text{p}}^H\mathbf{x}_{\text{p}}} \\ \notag
	 \text{s.t.} \quad&\Vert\mathbf{x}_{\text{p}}\Vert_2^2\leq \eta_0,
\end{align}
where $\eta_0$ denotes the transmit power budget.

Compared with the conventional approaches in~\cite{Wang2021WCL,Mohammadian2017CL}, which employ the average MIP as a surrogate objective to approximate problem ($\mathcal{P}1$), we directly address the original worst-case MIP minimization problem. To facilitate further problem transformation and algorithm design, we first introduce the following lemma.
\begin{lemma} \label{lemma1}
	The optimal solution of problem ($\mathcal{P}1$) is not unique. Specifically, if $\mathbf{x}_{\mathrm{opt}}$ is an optimal solution of ($\mathcal{P}1$), then any scaled vector $k\mathbf{x}_{\mathrm{opt}}$, with $k\in\mathbb{C}$ satisfying the power constraint, is also an optimal solution. Consequently, the set of optimal solutions forms an equivalence class.
\end{lemma}
\begin{proof}
	
	Assume that $\mathbf{x}_{\mathrm{opt}}$ attains the optimal value of problem ($\mathcal{P}1$). For any scalar $k\in\mathbb{C}$ such that $\Vert k\mathbf{x}_{\mathrm{opt}}\Vert_2^2 \leq \eta_0$, we have
	
	\begin{align} \label{proof}
		&\max_{1\leq i,j \leq M,i \neq j}\frac{(k{\mathbf{x}_{\mathrm{opt}}})^H\mathbf{H}_j^H\mathbf{H}_i(k{\mathbf{x}_{\mathrm{opt}}})(k{\mathbf{x}_{\mathrm{opt}}})^H\mathbf{H}_i^H\mathbf{H}_j(k{\mathbf{x}_{\mathrm{opt}}})}{(k{\mathbf{x}_{\mathrm{opt}}})^H(k{\mathbf{x}_{\mathrm{opt}}})(k{\mathbf{x}_{\mathrm{opt}}})^H(k{\mathbf{x}_{\mathrm{opt}}})} \\ \notag
		=&\max_{1\leq i,j \leq M,i \neq j}\frac{\vert k \vert^2}{\vert k \vert^2}\frac{\mathbf{x}_{\mathrm{opt}}^H\mathbf{H}_j^H\mathbf{H}_i\mathbf{x}_{\mathrm{opt}}\mathbf{x}_{\mathrm{opt}}^H\mathbf{H}_i^H\mathbf{H}_j\mathbf{x}_{\mathrm{opt}}}{\mathbf{x}_{\mathrm{opt}}^H\mathbf{x}_{\mathrm{opt}}\mathbf{x}_{\mathrm{opt}}^H\mathbf{x}_{\mathrm{opt}}}\\ \notag
		=&\max_{1\leq i,j \leq M,i \neq j}\frac{\mathbf{x}_{\mathrm{opt}}^H\mathbf{H}_j^H\mathbf{H}_i\mathbf{x}_{\mathrm{opt}}\mathbf{x}_{\mathrm{opt}}^H\mathbf{H}_i^H\mathbf{H}_j\mathbf{x}_{\mathrm{opt}}}{\mathbf{x}_{\mathrm{opt}}^H\mathbf{x}_{\mathrm{opt}}\mathbf{x}_{\mathrm{opt}}^H\mathbf{x}_{\mathrm{opt}}}
	\end{align}
	This shows that scaling $\mathbf{x}_{\mathrm{opt}}$ by any feasible complex scalar does not change the objective value of ($\mathcal{P}1$). 
	Hence, the optimal solution is not unique, and the proof is complete.
\end{proof}

%From \textbf{Lemma \ref{lemma1}} and its proof, it follows that the optimal solutions of problem ($\mathcal{P}1$) form an equivalence class due to the scale invariance of the objective function, and the transmit power of the pilot sequence does not affect the optimal value. Motivated by this observation, the optimization can be performed under a constant-modulus constraint without loss of optimality. Imposing a constant-modulus constraint renders the denominator of the objective function a constant, thereby simplifying the problem formulation. Consequently, without loss of generality, problem ($\mathcal{P}1$) can be equivalently transformed into the following normalized constant-modulus constrained optimization problem
According to \textbf{Lemma \ref{lemma1}}, the optimal solutions of problem ($\mathcal{P}1$) form an equivalence class due to the scale invariance of the objective function, implying that the pilot transmit power does not affect the optimum. Therefore, a constant-modulus constraint can be imposed without loss of optimality. Since this constraint renders the denominator constant, problem ($\mathcal{P}1$) can be equivalently reformulated as 
\begin{align} \label{P2}
	(\mathcal{P}2)\quad\min_{\mathbf{x}_{\text{p}}} \quad  &\max_{1\leq i,j \leq M,i \neq j}{\mathbf{x}_{\text{p}}^H\mathbf{H}_j^H\mathbf{H}_i\mathbf{x}_{\text{p}}\mathbf{x}_{\text{p}}^H\mathbf{H}_i^H\mathbf{H}_j\mathbf{x}_{\text{p}}} \\ \notag
	\text{s.t.} \quad&\Vert\mathbf{x}_{\text{p}}\Vert_2^2=1.
\end{align}

Considering the normalized constant-modulus constraint in problem ($\mathcal{P}2$), we employ the RCG method to solve it. To facilitate optimization, we first introduce the soft-max function to smoothly approximate the maximum operator~\cite{Song2025TCom}, given by
\begin{equation} \label{max_approx}
	\max_i\{a_i\}\approx u \log(\sum_{i}e^\frac{a_i}{u}),
\end{equation}
where $u>0$ is a tuning parameter satisfying 
\begin{equation} 
	\lim_{u\rightarrow0}u \log(\sum_{i}e^\frac{a_i}{u})=\max_i\{a_i\}.
\end{equation}

Using this approximation, problem ($\mathcal{P}2$) can be transformed into the following smooth optimization problem
\begin{align} \label{P3}
	(\mathcal{P}3)\quad\min_{\mathbf{x}_{\text{p}}} \quad  &u\log\left(\sum_{i=1}^{M}\sum_{j=1,j \neq i}^{M}e^{\frac{\mathbf{x}_{\text{p}}^H\mathbf{H}_j^H\mathbf{H}_i\mathbf{x}_{\text{p}}\mathbf{x}_{\text{p}}^H\mathbf{H}_i^H\mathbf{H}_j\mathbf{x}_{\text{p}}}{u}}\right) \\ \notag
	\text{s.t.} \quad&\Vert\mathbf{x}_{\text{p}}\Vert_2^2=1.
\end{align}

Let $\mathcal{M}=\{\mathbf{x}\in \mathbb{C}^{N\times 1}\vert \Vert\mathbf{x}\Vert_2^2=1\}$, which defines a complex unit sphere manifold in $\mathbb{C}^{N\times1}$. Accordingly, problem ($\mathcal{P}3$) can be equivalently reformulated as the following manifold-constrained optimization problem
\begin{align} \label{P4}
	(\mathcal{P}4)\quad\min_{\mathbf{x}\in \mathcal{M}}   \mu(\mathbf{x})=u\log\left(\sum_{i=1}^{M}\sum_{j=1,j \neq i}^{M}e^{\frac{\mathbf{x}^H\mathbf{H}_j^H\mathbf{H}_i\mathbf{x}\mathbf{x}^H\mathbf{H}_i^H\mathbf{H}_j\mathbf{x}}{u}}\right) .
\end{align}
In the following, we adopt Riemannian optimization techniques to efficiently solve problem ($\mathcal{P}4$) on the manifold $\mathcal{M}$.

\subsubsection{Calculation of the Riemannian Gradient} \label{step1}
The Riemannian gradient of the objective function $\mu(\mathbf{x})$ at the $n$-th iteration point $\mathbf{x}^{(n)} \in \mathcal{M}$ is obtained by projecting the Euclidean gradient onto the tangent space of the manifold $\mathcal{M}$ at $\mathbf{x}^{(n)}$, i.e.,
\begin{equation} \label{Riemann gradient}
	\nabla_{\mathcal{M}}\mu(\mathbf{x}^{(n)})=\text{P}_{\mathbf{x}^{(n)}}[\nabla\mu(\mathbf{x}^{(n)})],
\end{equation}
where $\mathrm{P}_{\mathbf{x}}(\cdot)$ denotes the projection operator at $\mathbf{x}$. For the complex unit-sphere manifold $\mathcal{M}$, the projection of an arbitrary vector $\boldsymbol{\alpha}$ onto the tangent space at $\mathbf{x}$ is given by
\begin{equation} \label{proj}
	\text{P}_{\mathbf{x}}(\bm{\alpha}) = \bm{\alpha}-\frac{1}{2}\mathbf{x}\left(\bm{\alpha}^H\mathbf{x}+\mathbf{x}^H\bm{\alpha}\right).
\end{equation}

Next, we derive the closed-form expression of the Euclidean gradient $\nabla\mu(\mathbf{x})$. Define
\begin{equation} 
	c_{ij}=\mathbf{x}^H\mathbf{H}_j^H\mathbf{H}_i\mathbf{x}, c_{ij}^*=\mathbf{x}^H\mathbf{H}_i^H\mathbf{H}_j\mathbf{x},
\end{equation}
and
\begin{equation} 
	g_{ij}(\mathbf{x})=\frac{\vert c_{ij}\vert^2}{u}=\frac{c_{ij}c_{ij}^*}{u}.
\end{equation} 
Then, the objective function can be expressed as
\begin{equation} 
	\mu(\mathbf{x})=u\log\left(\sum_{i=1}^{M}\sum_{j=1,j \neq i}^{M}e^{g_{ij}(\mathbf{x})}\right)=u\log\left[r(\mathbf{x})\right].
\end{equation}

By applying the chain rule, the Euclidean gradient of $\mu(\mathbf{x})$ is given by
\begin{align} \label{derivative1}
	\nabla\mu(\mathbf{x})&=\frac{u}{r(\mathbf{x})}\frac{\partial r(\mathbf{x})}{\partial \mathbf{x}}\\ \notag
	&=\frac{u}{r(\mathbf{x})}\left(\sum_{i=1}^{M}\sum_{j=1,j \neq i}^{M}e^{g_{ij}(\mathbf{x})}\frac{\partial g_{ij}(\mathbf{x})}{\partial \mathbf{x}}\right).
\end{align}
Furthermore, the derivative of $g_{ij}(\mathbf{x})$ with respect to $\mathbf{x}$ can be expressed as
\begin{align} \label{derivative2}
	\frac{\partial g_{ij}(\mathbf{x})}{\partial \mathbf{x}}=\frac{1}{u}\left(\frac{\partial c_{ij}}{\mathbf{x}}c_{ij}^*+c_{ij}\frac{\partial c_{ij}^*}{\mathbf{x}}\right),
\end{align}
where 
\begin{align} \label{derivative3}
	\frac{\partial c_{ij}}{\mathbf{x}}=\frac{\partial c_{ij}^*}{\mathbf{x}}=\mathbf{H}_j^H\mathbf{H}_i\mathbf{x}+\mathbf{H}_i^H\mathbf{H}_j\mathbf{x}.
\end{align}

Substituting (\ref{derivative2}) and (\ref{derivative3}) into (\ref{derivative1}), the Euclidean gradient $\nabla\mu(\mathbf{x})$ can be obtained in closed form, which is shown at the top of this page.
\begin{figure*} \label{RGD}
	\begin{align} \label{Euclidean gradient}
		\nabla\mu(\mathbf{x})=\frac{\sum_{i=1}^{M}\sum_{j=1,j \neq i}^{M}e^{\frac{1}{u}\mathbf{x}^H\mathbf{H}_j^H\mathbf{H}_i\mathbf{x}\mathbf{x}^H\mathbf{H}_i^H\mathbf{H}_j\mathbf{x}}\left(\mathbf{x}^H\mathbf{H}_i^H\mathbf{H}_j\mathbf{x}+\mathbf{x}^H\mathbf{H}_j^H\mathbf{H}_i\mathbf{x}\right)\left(\mathbf{H}_j^H\mathbf{H}_i+\mathbf{H}_i^H\mathbf{H}_j\right)}{\sum_{i=1}^{M}\sum_{j=1,j \neq i}^{M}e^{\frac{1}{u}\mathbf{x}^H\mathbf{H}_j^H\mathbf{H}_i\mathbf{x}\mathbf{x}^H\mathbf{H}_i^H\mathbf{H}_j\mathbf{x}}}\mathbf{x}.
	\end{align}
\end{figure*}

\subsubsection{Update of the Descent Direction and Step Size on the Tangent Space}

Let $\mathbf{b}^{(n)}$ denote the descent direction at the $n$-th iteration. In the RCG method, $\mathbf{b}^{(n)}$ is determined by the current Riemannian gradient $\nabla_{\mathcal{M}}\mu(\mathbf{x}^{(n)})$ and the projection of the previous descent direction $\mathbf{b}^{(n-1)}$ onto the tangent space at $\mathbf{x}^{(n)}$, i.e.,
\begin{equation} \label{direction}
	\mathbf{b}^{(n)}=-\nabla_{\mathcal{M}}\mu(\mathbf{x}^{(n)})+\beta^{(n)}\text{P}_{\mathbf{x}^{(n)}}\left(\mathbf{b}^{(n-1)}\right),
\end{equation}
where 
\begin{align} \label{step size}
	\beta^{(n)}&=\\ \notag
	&\frac{\left[\nabla_{\mathcal{M}}\mu\left(\mathbf{x}^{(n)}\right)\right]^H\left[\nabla_{\mathcal{M}}\mu\left(\mathbf{x}^{n}\right)-\text{P}_{\mathbf{x}^{(n)}}\left(\nabla_{\mathcal{M}}\mu\left(\mathbf{x}^{(n-1)}\right)\right)\right]}{\Vert\nabla_{\mathcal{M}}\mu\left(\mathbf{x}^{(n-1)}\right)\Vert_2^2},
\end{align}

Given the descent direction $\mathbf{b}^{(n)}$, the intermediate update on the tangent space is performed as
\begin{equation} \label{point update1}
	\tilde{\mathbf{x}}^{(n+1)}=\mathbf{x}^{(n)}+\delta^{(n)}\mathbf{b}^{(n)},
\end{equation}
where $\delta^{(n)}$ denotes the step size, which is determined via a backtracking line search based on the Armijo rule.
\subsubsection{Retraction onto the Complex Unit-Sphere Manifold}  \label{step3}

After the tangent-space update, the point $\tilde{\mathbf{x}}^{(n+1)}$ is generally no longer on the manifold $\mathcal{M}$. Therefore, a retraction operation is applied to project it back onto the complex unit-sphere manifold, yielding
\begin{equation} \label{point update2}
	{\mathbf{x}}^{(n+1)} = \frac{\tilde{\mathbf{x}}^{(n+1)}}{\Vert \tilde{\mathbf{x}}^{(n+1)} \Vert_2}.
\end{equation}

By iteratively executing steps (1)-(3), the solution $\mathbf{x}_0$ of problem ($\mathcal{P}4$) can be obtained. Accordingly, the optimal solution of the original problem ($\mathcal{P}1$) is recovered as
\begin{equation} \label{recovery}
	\mathbf{x}^*=\sqrt{\eta_0}\mathbf{x}_0.
\end{equation}

The overall RCG-based algorithm for solving problem $(\mathcal{P}1)$ is summarized in Algorithm~\ref{alg:RCG1}. According to~\cite{Song2025TCom}, the computational complexity of Algorithm~\ref{alg:RCG1} is $\mathcal{O}\left(N_m M\right)$, where $N_m$ denotes the maximum number of iterations.

\renewcommand{\algorithmicrequire}{\textbf{Input:}}% Input
\renewcommand{\algorithmicensure}{\textbf{Output:}}% Output
\begin{algorithm}[t]
	\caption{RCG-based AF-domain pilot optimization}
	\label{alg:RCG1}
	\begin{algorithmic}[1]
		\Require Approximation parameter $u$, transmit power $\eta_0$, channel matrices $\{\mathbf{H}_i\}_{i=0}^{M-1}$, convergence threshold $\epsilon$, and maximum number of iterations $m_\text{max}$ and $N_m$.
		\State Initialize $\mathbf{x}^{(0)} \in \mathcal{M}$ and $\mathbf{b}^{(0)} = \mathbf{0}$;
		\State Select line search parameters $\lambda \in (0,1)$ and $\zeta \in (0,0.5)$;
		\State Set iteration index $n = 1$ and initialize $\kappa = \infty$;
		
		\While {$\kappa > \epsilon$ and $n \leq N_{\text{max}}$}
		\State Compute the Riemannian gradient $\nabla_{\mathcal{M}}\mu(\mathbf{x}^{(n)})$ based on~(\ref{Riemann gradient}), (\ref{proj}), and (\ref{Euclidean gradient});
		\State Compute the descent direction $\mathbf{b}^{(n)}$ based on~(\ref{direction}) and (\ref{step size});
		\State Set $m = 0$;
		\For{ $0\leq m \leq m_\text{max}$}
		\If{$\mu(\mathbf{x}^{(n)}+\lambda^m\mathbf{b}^{(n)})\leq \zeta \lambda^m \nabla^T\mu(\mathbf{x}^{(n)})\mathbf{b}^{(n)}+\mu(\mathbf{x}^{(n)})$}
		\State{Break; }
		\Else
		\State{$m\leftarrow m+1$;}
		\EndIf
		\EndFor
		\State Set step size $\delta^{(n)} = \lambda^m$;
		\State Update $\tilde{\mathbf{x}}^{(n+1)}$ based on~(\ref{point update1});
		\State Update ${\mathbf{x}}^{(n+1)}$ based on~(\ref{point update2});
		\State Increment $n \leftarrow n + 1$.
		\State Compute $\kappa = \Vert \mathbf{x}^{(n+1)} - \mathbf{x}^{(n)} \Vert_2^2$.
		\EndWhile
		\State Recover transmit power to obtain $\mathbf{x}^*$ based on~(\ref{recovery}).
		\Ensure Optimized pilot sequence $\mathbf{x}^*$.
	\end{algorithmic}
\end{algorithm}

\begin{rmk}
	It is also worth noting that the proposed pilot optimization algorithm can be executed offline; therefore, its computational complexity does not impose additional burden on practical communication systems.
\end{rmk}

\section{TCN-based Received Pilots Prediction}
\label{Receive sequence prediction}
In this section, we investigate the design of the pilot prediction module at the receiver, as illustrated in Fig.~\ref{fig:sys}. We analyze the inherent AR structure of the received pilot symbol sequence and then propose a TCN-based pilot prediction framework. 

\subsection{AR Property of Received Pilot Symbol Sequence}
Under the assumption that the channel delay and Doppler parameters remain approximately constant within each OFDM slot, we consider that the transmitter sends the same pilot sequence $\mathbf{x}_{\text{p}}$ on each OFDM symbol. To characterize the relationship among the received pilot symbols, we further assume a noise-free propagation for analytical tractability. After propagating through the $P$-path channel described in (\ref{CIR}), the received AF-domain pilot corresponding to the $k$-th symbol can be expressed as
\begin{equation} \label{ypk}
	\mathbf{y}_{\text{p},k}=\mathbf{H}_{\text{AFD},k}\mathbf{x}_{\text{p}}=\sum_{i=0}^{P-1}\tilde{h}_{i,k}\mathbf{H}_{i}\mathbf{x}_{\text{p}},
\end{equation}
where $\mathbf{x}_{\text{p}}$ denotes the designed pilot sequence obtained in the previous section. Based on (\ref{htilde}), the $n$-th element of $k$-th received pilot symbols can be expressed as
\begin{align}\label{ypkn}
	{y}_{\text{p},k,n}=\sum_{i=0}^{P-1}\tilde{h}_{i,k}\mathbf{H}_{i}(n,:)\mathbf{x}_{\text{p}}
	%&=\sum_{i=0}^{P-1}\tilde{h}_{i,0}e^{-j2\pi \nu_i(N+L)k/N}\mathbf{H}_{i}(n,:)\mathbf{x}_{\text{p}}
	\triangleq\sum_{i=0}^{P-1}c_{n,i}e^{-j\theta_ik},
\end{align}
where $\mathbf{H}_{i}(n,:)$ denotes the $n$-th row vector of $\mathbf{H}_{i}$, $c_{n,i}=\tilde{h}_{i,0}\mathbf{H}_{i}(n,:)\mathbf{x}_{\text{p}}$, and $\theta_i=2\pi \nu_i(N+L_\text{CP})/N$. 

As shown in (\ref{ypkn}), for any fixed index $n$, the sequence $\{{y}_{\text{p},k,n}\}(k=0,1,\ldots)$ can be expressed as a linear combination of $P$ complex exponential terms $\{e^{-j\theta_ik}\}(i=0,1,\ldots,P-1)$. It is a well-established principle that any sequence composed of a finite sum of exponential terms satisfies a $P$-th order homogeneous linear difference equation with constant complex coefficients, whose characteristic roots are given by $\{e^{-j\theta_i}\}(i=0,1,\ldots,P-1)$. Consequently, there exists a set of constant complex coefficients $\{\beta_0,\ldots,\beta_{P-1}\}$ such that
\begin{equation}\label{nDQ}
	y_{\text{p},k+P,n} + \beta_{P-1}y_{\text{p},k+P-1,n} + \cdots + \beta_0 y_{\text{p},k,n} = 0,\quad \forall k,
\end{equation}

which implies that sample $y_{\text{p},k+P,n}$ can be linearly predicted from the past $P$ samples $\{y_{\text{p},k,n},\dots,y_{\text{p},k+P-1,n}\}$.

It is further observed that each component of $\mathbf{y}_{\text{p},k}$ satisfies a difference equation of the same form. Moreover, since the characteristic roots associated with all components are identical, i.e., $\{e^{-j\theta_i}\}(i=0,1,\ldots,P-1)$, they share a common set of prediction coefficients. As a result, the received AF-domain pilot symbol satisfies the following unified $P$-th order vector linear difference equation
\begin{equation}\label{vDQ}
	\mathbf{y}_{\text{p},k+P} + \beta_{P-1}\mathbf{y}_{\text{p},k+P-1} + \cdots + \beta_0 \mathbf{y}_{\text{p},k} = \mathbf{0},\quad \forall k,
\end{equation}

\begin{equation}\label{11}
	\mathcal{O}\left(N^3\right),
\end{equation}

which can be equivalently written in the AR form as
\begin{equation}\label{yAR}
	\mathbf{y}_{\text{p},k+P}=\gamma_{0}\mathbf{y}_{\text{p},k} + \cdots + \gamma_{P-1} \mathbf{y}_{\text{p},k+P-1},\quad \forall k.
\end{equation}

Based on (\ref{yAR}), the received AF-domain pilot symbol sequence exhibits an inherent AR structure, which provides the theoretical foundation for the proposed inference of virtual pilot symbols. Building upon the above analysis, we present the following conclusion, formally stated as \textbf{Lemma~\ref{lemma2}}.
\begin{lemma} \label{lemma2}
	Within the interval over which the channel delay and Doppler parameters remain approximately constant, repeated transmission of the same pilot sequence $\mathbf{x}_{\text{p}}$ leads to a received AF-domain pilot sequence $\{\mathbf{y}_{\text{p},k}\}$ that satisfies the $P$-th order AR model in (\ref{yAR}).
\end{lemma}
Similarly, \textbf{Lemma~\ref{lemma2}} can be extended to both time-frequency (TF) domain and DD domain. Moreover, the AR property is not restricted to the received pilot sequence; the underlying channel matrix itself also exhibits a similar AR structure. In essence, the $P$-th order AR behavior of the received pilots is a direct consequence of the inherent AR property of the channel matrix, leading to the following \textbf{Theorem~\ref{theorem1}}.
\begin{thm} \label{theorem1}
	Within the interval over which the channel delay and Doppler parameters remain approximately constant, the channel matrix $\{\mathbf{H}_{k}\}$ possesses the $P$-order AR property, which can be expressed as 
	\begin{equation}\label{HAR}
		\mathbf{H}_{k+P}=\gamma_{0}\mathbf{H}_{k} + \cdots + \gamma_{P-1} \mathbf{H}_{k+P-1},\quad \forall k.
	\end{equation}
\end{thm}

In static or low-Doppler scenarios, interpolation between observed pilots can approximately capture the relationship described in (\ref{HAR}). However, as the Doppler shift increases, interpolation becomes ineffective, leading to significant performance degradation. 
Motivated by (\ref{HAR}), the future CSI can instead be predicted using the most recent $P$ channel matrices. Specifically, the AR coefficients $\gamma_{k}$ $(k = 0, \ldots, P-1)$ are first estimated from the previous $P+1$ received pilot symbols using the least-squares (LS) method, expressed as
\begin{equation}\label{LS}
	\bm{\gamma}=\left(\mathbf{Y}\mathbf{Y}^H\right)^{-1}\mathbf{Y}^H\mathbf{y}_{k+P},
\end{equation}
where $\bm{\gamma} = [\gamma_0, \ldots, \gamma_{P-1}]^T$, and $\mathbf{Y} = [\mathbf{y}_{k}, \ldots, \mathbf{y}_{k+P-1}]$. Once $\bm{\gamma}$ is obtained, the subsequent channel matrices can be predicted using (\ref{HAR}). %However, this classic and direct approach is inherently unstable due to the following challenges:
%\begin{itemize}
%	\item[$\bullet$] The received pilots are corrupted by unpredictable noise, resulting in unreliable estimation of AR coefficients $\bm{\gamma}$.
%	\item[$\bullet$] The observed CSI is obtained from channel estimation, which is subject to inevitable estimation errors. As a result, the ideal AR relationship in (\ref{HAR}) is often poorly satisfied in practice.
%	%\item[$\bullet$] To maintain communication efficiency, it is necessary to predict as long CSI as possible based on a limited number of known CSI samples. This requires iterative prediction, during which prediction errors accumulate over time, resulting in progressively degraded accuracy for later CSI predictions.
%\end{itemize}
%\begin{rmk}
%	Even under ideal noise-free conditions with perfect CSI, numerical simulations show that the difference equations derived from the above method are unstable according to the Von Neumann criterion~\cite{charney1950numerical}. This lack of numerical stability indicates that the approach is not robust and, therefore, cannot be considered a reliable algorithm. Nonetheless, \textbf{Lemma~\ref{lemma2}} still uncovers an inherent relationship among adjacent channel matrices, which motivates us to adopt a data-driven prediction strategy using deep learning (DL).
%\end{rmk}
However, the direct AR-based predictor is inherently unstable in practice due to noisy pilot observations and channel estimation errors, which prevent the ideal AR relationship in (\ref{HAR}) from being accurately satisfied. Furthermore, even with perfect CSI and noise-free observations, our numerical simulations show that the resulting difference equations fail the Von Neumann stability criterion~\cite{charney1950numerical}. Nevertheless, \textbf{Lemma~\ref{lemma2}} reveals an inherent temporal correlation among adjacent channel matrices, which motivates the development of a data-driven deep learning (DL)-based predictor.

\subsection{TCN-based Pilot Prediction Network Design}
In contrast to the LSTM-based pilot predictor in~\cite{tang2026dopplerresilientleosatelliteofdm}, which restricts pilots to the beginning of each slot, practical systems often require more flexible pilot placements. To accommodate this, we reformulate pilot prediction as a task analogous to single image super-resolution (SISR)~\cite{Li2020WCL}, where the observed pilots are treated as a low-resolution input and the complete pilot sequence, including both observed and virtual pilots, is reconstructed as the high-resolution output.

\begin{figure*}[t]
	\centering
	\includegraphics[width=0.85\textwidth]{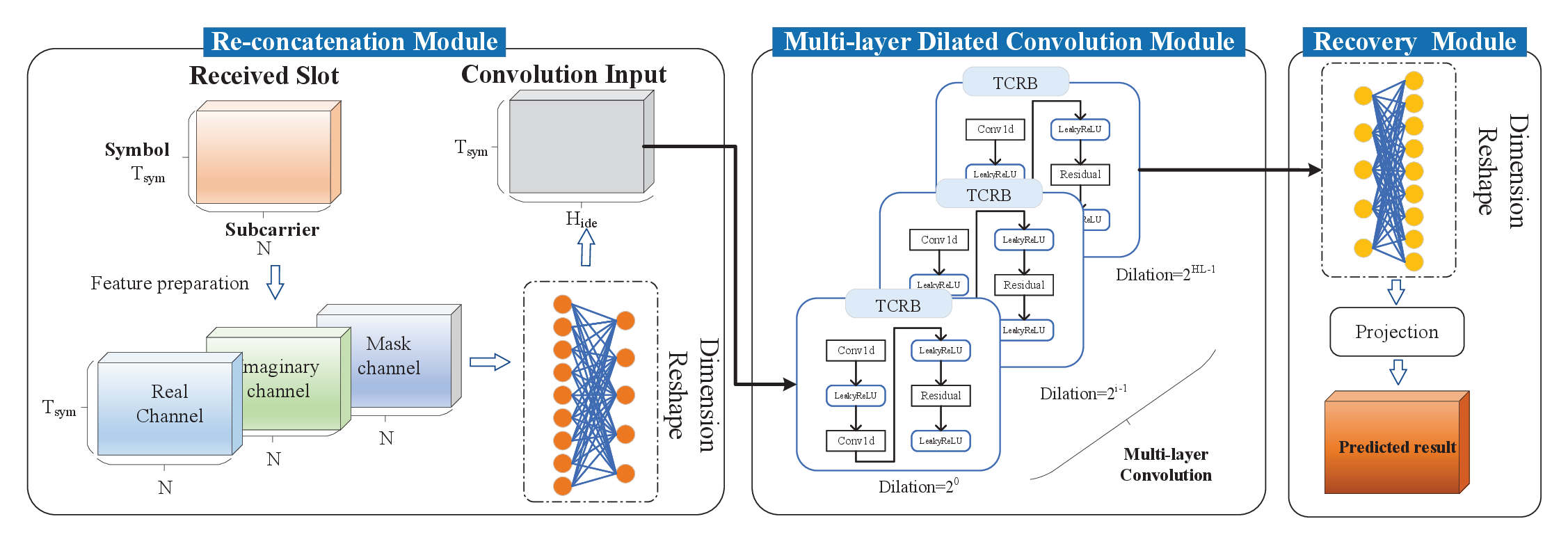}
	\caption{Structure of TCN-based prediction network.}
	\label{fig:nnTCN}
\end{figure*}

Based on this formulation, convolutional neural networks (CNNs) are employed to exploit local structural features. Furthermore, leveraging \textbf{Lemma~\ref{lemma2}} to capture temporal dependencies, we adopt a temporal convolutional network, and the detailed architecture is illustrated in Fig.~\ref{fig:nnTCN}. Specifically, it can be seen that the TCN-based prediction NN consists of three key modules: a \textbf{Re-concatenation Module}, a \textbf{Multi-layer Dilated Convolution Module}, and a \textbf{Recovery Module}.

\subsubsection{Re-concatenation Module}
During each training epoch, the input to the network $\mathbf{Y}_{\text{p}} \in \mathbb{C}^{T_{\text{sym}} \times N}$, where $T_{\text{sym}}$ denotes the number of symbols per slot, is a partially observed complex-valued pilot matrix. Specifically, $\mathbf{Y}_{\text{p}}$ retains pilot symbols only at predefined pilot positions, while all remaining entries are set to zero.
To facilitate NN processing, the complex-valued matrix is decomposed into its real and imaginary components. In addition, a binary mask is introduced to explicitly indicate the locations of the observed pilot symbols, which provides auxiliary structural information to the network. These three components are then stacked to form a $3N$-channel input matrix.
This transformation aligns all subcarrier information at each symbol index within the convolutional channel dimension, enabling the convolutional kernels along the time dimension to simultaneously capture the full subcarrier state. As a result, the neural network can effectively learn the AR properties along the time dimension. The corresponding operation can be represented as
\begin{equation}\label{input_TCN}
	\mathbf{A}_{\text{p}}= [\operatorname{Re}\left(\mathbf{Y}_{\text{p}}\right), \operatorname{Im}\left(\mathbf{Y}_{\text{p}}\right), \mathbb{I}\left(\mathbf{Y}_{\text{p}}\right)]\in \mathbb{R}^{ T_{\text{sym}} \times 3N},
\end{equation}
where $\mathbb{I}\left(\cdot\right)$ denotes the indicator function.

Then, a point-wise convolution is applied to project the input into a lower-dimensional feature space. This operation performs channel-wise linear transformations, allowing the model to aggregate subcarrier-wise features together with mask information and to learn more expressive and compact representations. The corresponding operation can be expressed as
\begin{equation}\label{linear1}
	\mathbf{G}^{(0)}= \mathcal{L}_1(\mathbf{A}_{\text{p}})\in \mathbb{R}^{T_{\text{sym}} \times H_\text{ide}},
\end{equation}
where $H_{\text{ide}}$ denotes the feature dimension of the input to the subsequent TCN-based network, and $\mathcal{L}_1(\cdot)$ represents the input linear transformation implemented by the point-wise convolution.
\subsubsection{Multi-layer Dilated Convolution Module}
The core functional unit of the proposed TCN-based neural network is the temporal convolutional and residual block (TCRB). To effectively capture the time-varying characteristics of high-mobility channels across symbols, each TCRB integrates dilated non-causal convolutions with residual learning mechanisms. Unlike causal TCNs, symmetric padding is adopted in this work, allowing the model to exploit both past and future contextual information, thereby improving reconstruction accuracy.

To capture long-range dependencies along the time dimension, dilated convolutions are employed. Unlike standard convolutions, dilated convolutions introduce gaps between kernel elements, controlled by the dilation factor $d$, which enables the network to cover a wider temporal range without increasing the number of parameters. In the proposed architecture, the dilation factor increases exponentially with the network depth. As a result, the shallow layers focus on local temporal patterns, while deeper layers progressively capture broader temporal dependencies, enabling efficient modeling of both short-term and long-term variations.

To mitigate the vanishing gradient problem inherent in deep architectures, a residual connection is incorporated within each TCRB. Specifically, the output of the $i$-th TCRB denoted as $\mathbf{G}^{(i)}$ is given by
\begin{equation}\label{residual}
	\mathbf{G}^{(i)} = \sigma \big( \mathcal{F}^{(i)}(\mathbf{G}^{(i-1)}) + \mathcal{R}(\mathbf{G}^{(i-1)}) \big),
\end{equation}
where $\mathcal{F}^{(i)}(\cdot)$ represents the nonlinear transformation implemented by the stacked dilated convolutional layers in the $i$-th TCRB, $\mathcal{R}(\cdot)$ denotes the residual mapping, and $\sigma(\cdot)$ is the LeakyReLU activation function.
\subsubsection{Recovery Module} After temporal feature extraction, another point-wise convolution is used to map the hidden features back to the original signal space, and the corresponding operation can be expressed as
\begin{equation}\label{linear2}
	\mathbf{B_\text{p}}= \mathcal{L}_2(\mathbf{G}^{(\text{HL})})\in \mathbb{R}^{T_{\text{sym}} \times 2N},
\end{equation}
where $\text{HL}$ denotes the layer number of multiple-TCRBs, and $\mathcal{L}_2(\cdot)$ represents the output linear transformation implemented by the point-wise convolution. 

Then, the prediction result $\mathbf{B_\text{p}}$ is reshaped to reconstruct the complex-valued pilot symbols from the real-valued outputs, which can be expressed as
\begin{equation}\label{reconstruct1}
	\tilde{\mathbf{Y}}_{\text{p}}= \mathbf{B_\text{p}}(:, 0:N-1)+1j\cdot\mathbf{B_\text{p}}(:, N:2N-1).
\end{equation}

%To ensure that the reconstructed output remains consistent with the known pilot observations, a data consistency mechanism is incorporated. Specifically, the network prediction is combined with the original input via the binary mask, such that the observed pilot entries are preserved while only the missing entries are estimated. The corresponding operation can be formulated as
To ensure consistency with the known pilot observations, a data consistency mechanism is introduced. The network output is combined with the original input via a binary mask, preserving the observed pilot entries while estimating only the missing ones. The operation can be formulated as follows
\begin{equation}\label{msk}
	\hat{\mathbf{Y}}_{\text{p}}= \tilde{\mathbf{Y}}_{\text{p}} \odot \left(\mathbf{1}_{T_{\text{sym}} \times N}-\mathbb{I}\left(\mathbf{Y}_{\text{p}}\right)\right)+\mathbf{Y}_{\text{p}} \odot \mathbb{I}\left(\mathbf{Y}_{\text{p}}\right),
\end{equation}
where $\hat{\mathbf{Y}}_{\text{p}}$ denotes the predicted result, and $\odot$ denotes the Hadamard product.
\subsubsection{Training Strategy}
The proposed predictor is trained offline using simulated data, where $\mathbf{Y}_{\text{p}}$ are generated under diverse channel conditions to enhance generalization capability. The network parameters $\bm{\Theta}$ are optimized according to the mean square error (MSE) criterion, and the corresponding loss function is defined as
\begin{equation}\label{loss2}
	J(\bm{\Theta})= \mathbb{E}\{\Vert{\hat{\mathbf{Y}}_{\text{p}}} - \mathbf{Y}_{\text{p},\text{T}}\Vert_F^2\},
\end{equation}
where $\mathbf{Y}_{\text{p},\text{T}}$ denotes the ground-truth pilot matrix.

\section{Simulation Results}
\label{Simulation Results}
In this section, the effectiveness of the proposed pilot optimization scheme and pilot prediction networks is respectively evaluated via Monte Carlo simulations. To reflect realistic conditions, the 3GPP time delay line (TDL)-D channel model~\cite{3gpp.38.811} is adopted, with a maximum receiver speed of \textcolor{black}{100} km/h, while the remaining simulation parameters are summarized in Table~\ref{tab:sim_params}. The simulation is performed over more than 2000 slots, and for the AF-domain channel estimation, the OMP-based estimator and the SBL-based estimator presented in~\cite{Tang2025ICCC} are employed. The minimum mean square error (MMSE) equalizer is employed for the data detection. 

\begin{table}[t]
	\caption{Simulation Parameters}
	\label{tab:sim_params}
	\centering
	\begin{tabular}{|c|c|c|}
		\hline
		\textbf{Parameter} & \textbf{Symbol} &\textbf{Value} \\
		\hline
		Subcarrier number & $N$ &  $64$ \\
		\hline
		symbol number per slot & $T_\text{sym}$ & 12\\
		\hline
		Carrier frequency & $f_c$ &  \textcolor{black}{5.8}~\text{GHz} \\
		\hline
		Subcarrier spacing & $\Delta f$ & 15~\text{kHz}\\
		\hline
		\multirow{2}{*}{Chirp parameter} 
		& $c_1$ & $0.0391$ \\
		\cline{2-3}
		& $c_2$ & $0$ \\
		\hline
		\textcolor{black}{Modulation order} & $Q$ & 4 \\
		\hline
	\end{tabular}
\end{table}
\textcolor{black}{We adopt the LSTM-based predictor with two pilot configurations proposed in~\cite{tang2026dopplerresilientleosatelliteofdm} as the baseline. For the LSTM-based scheme, pilot symbols are placed at the beginning of each slot (i.e., at the 1st–2nd and 1st–4th symbol positions), whereas the proposed TCN-based predictor inserts pilots at the 3rd and 7th symbols within each slot (symbols 0–11).} The training dataset consists of 6000 slots containing full pilots, generated at an SNR of 25 dB. MSE is adopted as the loss function, and the Adam optimizer is employed with an initial learning rate of $10^{-3}$. The learning rate is gradually decreased following a cosine annealing schedule throughout the training process.

\subsection{AF-domain Pilot and its Optimization Scheme Evaluation}
In this subsection, we present simulation results to evaluate the performance of the proposed AF-domain pilot and its RCG-based optimization scheme, respectively. To focus exclusively on the effectiveness of the AF-domain pilot design, the interpolation error is assumed to be perfectly compensated, while the impact of interpolation errors will be investigated in the next subsection.
\begin{figure}[H]
	\centering
	\includegraphics[width=0.45\textwidth]{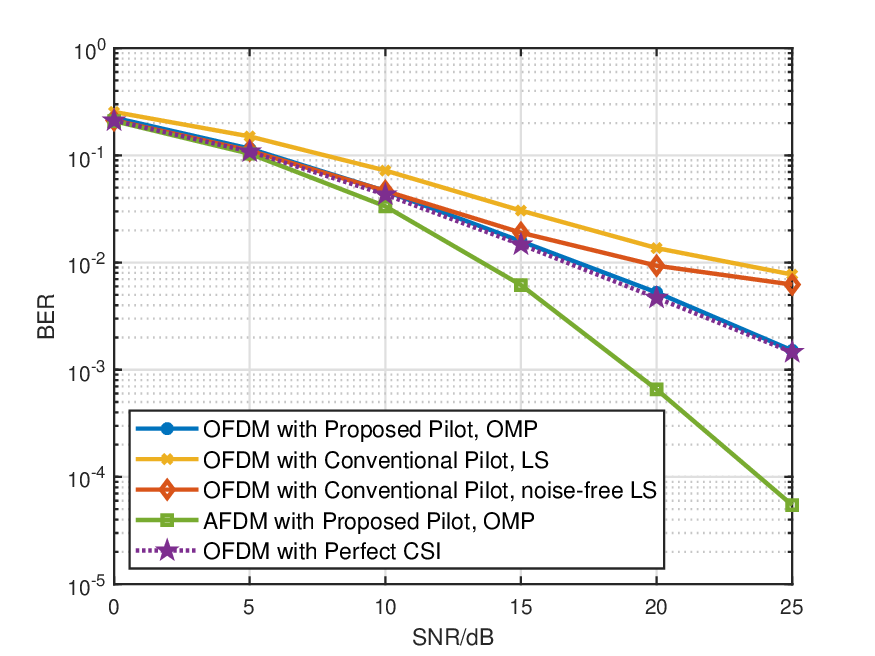}
	\caption{The BER performance comparison between AF-domain and TF-domain pilot configurations.}\label{fig:1}
\end{figure}
To demonstrate the effectiveness of the proposed AF-domain pilot, the BER performance comparison between the conventional TF-domain pilot and the proposed AF-domain pilot is first illustrated in Fig.~\ref{fig:1}. In this figure, ``noise-free LS" serves as a benchmark, where the pilot SNR is assumed to be infinite and the CP-based Doppler shift estimation and compensation scheme in~\cite{Nishad2013CICT} is employed at the receiver. As shown in Fig.~\ref{fig:1}, the proposed AF-domain pilot significantly outperforms the conventional TF-domain pilot, even when the latter operates under an infinite pilot SNR. This result indicates that the performance bottleneck of OFDM systems in high-Doppler scenarios is mainly originate from the ICI induced by Doppler spread. Specifically, conventional OFDM pilots only estimate the diagonal elements of the frequency-domain channel matrix and therefore fail to capture the off-diagonal ICI components under high-mobility conditions. Even when Doppler compensation techniques are employed, their effectiveness remains limited due to the Doppler spread caused by multipath components with various Doppler shifts.
In contrast, the proposed AF-domain pilot can effectively estimate the Doppler effects of individual propagation paths, thereby avoiding the BER error floor observed in conventional pilot schemes. Owing to its high channel estimation accuracy, the proposed pilot achieves BER performance close to the perfect CSI benchmark, demonstrating strong performance and robustness under high-Doppler conditions. \textcolor{black}{Finally, it can be observed that the AFDM scheme achieves superior BER performance compared with OFDM under the same pilot configuration, as AFDM data can obtain full multipath diversity that is not available in OFDM systems~\cite{Bemani2023TWC}. Considering the resource allocation flexibility and low-complexity processing of OFDM-based data transmission and reception, this highlights a fundamental trade-off among diversity gain, implementation complexity, and compatibility with existing OFDM systems.}

To assess the effectiveness of the proposed pilot optimization design, we then compare the channel estimation performance under different AF-domain pilot configurations. Specifically, the normalized mean square error (NMSE) is adopted as the performance metric, which is defined as
%, since a well-designed pilot can reduce the mutual coherence of the resulting  dictionary according to (\ref{MIP1}), thereby enhancing dictionary identifiability and improving channel estimation accuracy. 
\begin{equation} 
	\text{NMSE}  = 10\log_{10}\mathbb{E}\left\{\frac{\Vert\tilde{\mathbf{H}}_{\text{es}}-{\mathbf{H}}_{\text{true}}\Vert_F^2}{\Vert{\mathbf{H}}_{\text{true}}\Vert_F^2}\right\},
\end{equation}
where $\tilde{\mathbf{H}}_{\text{es}}$ denotes the estimated effective channel matrix, and $\mathbf{H}_{\text{true}}$ denotes the true effective channel matrix.

\begin{figure}
	\centering
	\includegraphics[width=0.45\textwidth]{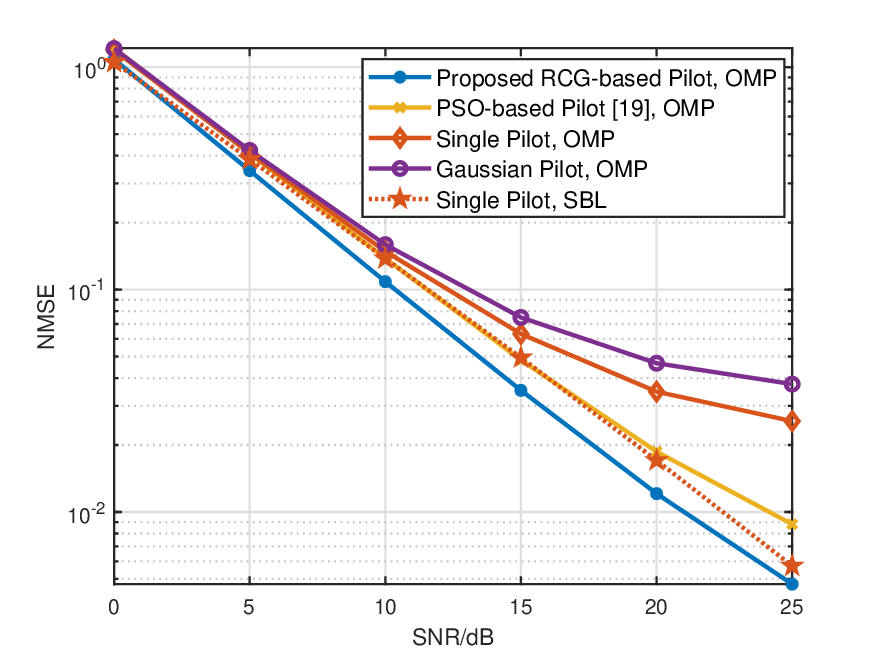}
	\caption{The NMSE performance comparison among different pilot configurations and estimators.}\label{fig:2}
\end{figure}
The NMSE performance under different AF-domain pilot configurations and estimators is first illustrated in Fig.~\ref{fig:2}. In addition to the conventional single-pilot and Gaussian random pilot schemes, the particle swarm optimization (PSO)-based pilot design in~\cite{Wang2021WCL} is also included for comparison, where the algorithm parameters are chosen according to the recommendations in~\cite{Wang2021WCL}. As shown in Fig.~\ref{fig:2}, both the single-pilot and Gaussian random pilot schemes combined with OMP-based estimator suffer from a pronounced error floor in the high-SNR region. This is mainly because the dictionaries generated by these pilots exhibit relatively high mutual coherence, which degrades sparse recovery capability and limits the performance of the CS-based channel estimator. Although the single-pilot scheme outperforms the Gaussian random pilot, indicating that a structured pilot design can partially improve estimation performance, it still fails to eliminate the error floor, highlighting the necessity of pilot optimization. By contrast, both optimized pilot schemes significantly improve the NMSE performance by effectively reducing dictionary coherence. 

Moreover, the proposed RCG-based pilot optimization consistently outperforms the PSO-based scheme over the entire SNR range. This performance gain is attributed to the fact that the PSO-based method minimizes an averaged MIP metric, which only approximately characterizes the worst-case column correlation in (\ref{MIP1}). In contrast, the proposed RCG-based optimization directly targets the maximum inner product criterion, resulting in a dictionary with lower worst-case coherence. Consequently, the sparse channel components become more distinguishable, enabling more reliable support recovery and more accurate channel estimation. On the other hand, it can be observed that the single-pilot scheme achieves a significant NMSE performance improvement when the SBL-based estimator is employed. This gain is mainly attributed to the fact that the SBL-based estimator is less sensitive to the strong column coherence of the sensing dictionary, at the expense of increased computational complexity. Nevertheless, the proposed RCG-based pilot combined with the low-complexity OMP-based estimator still outperforms the single-pilot scheme with the SBL-based estimator, demonstrating that the proposed pilot design can achieve superior channel estimation performance without introducing additional computational complexity at the receiver.

\begin{figure}[H]
	\centering
	\includegraphics[width=0.45\textwidth]{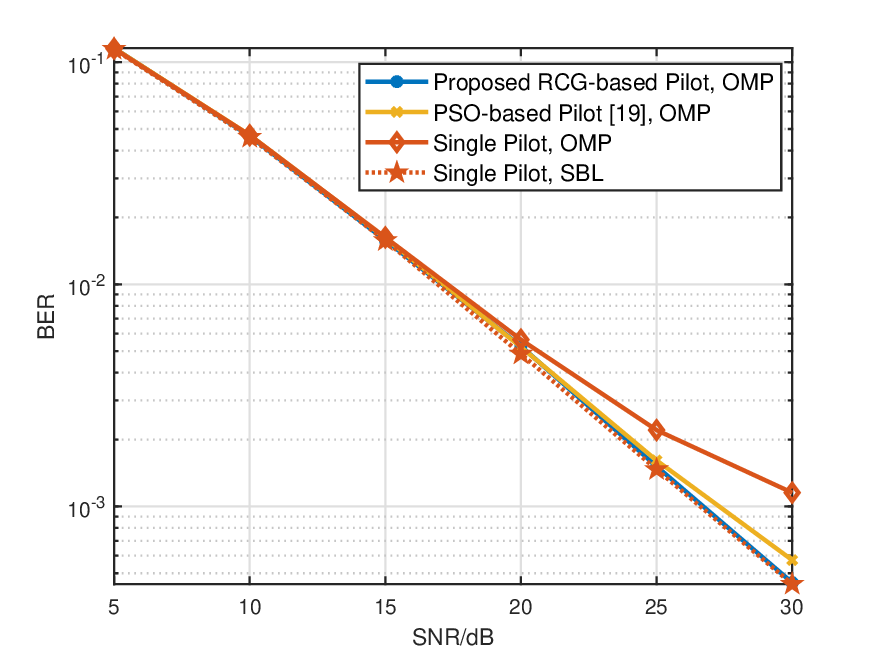}
	\caption{The BER performance comparison among different pilot configurations and estimators.}\label{fig:3}
\end{figure}
We then evaluate the BER performance under different AF-domain pilot configurations and estimators as shown in Fig.~\ref{fig:3}. As illustrated, the proposed RCG-based pilot combined with the OMP estimator still achieves the best BER performance over the entire SNR range, demonstrating the effectiveness of the proposed pilot optimization in improving sensing dictionary quality and channel estimation accuracy. In contrast, the conventional single-pilot scheme with the OMP estimator exhibits inferior BER performance due to the strong column coherence of the resulting estimation dictionary. Although the PSO-based pilot improves the BER performance, a noticeable performance gap still exists compared with the proposed RCG-based pilot scheme in the high SNR region. By employing the SBL-based estimator, the BER performance of the single-pilot scheme is significantly improved, although at the cost of higher computational complexity. It is worth noting that the proposed RCG-based pilot with the low-complexity OMP estimator achieves BER performance comparable to that of the single-pilot scheme with the SBL-based estimator, indicating that effective pilot optimization can provide substantial performance gains without increasing the receiver complexity.

\subsection{Performance of the Proposed Predictors Evaluation}
In this subsection, we present simulation results to evaluate the effectiveness of the proposed predictors at the receiver.
The detailed parameter configurations of the compared and proposed predictors are summarized in Tables~\ref{tab:net_params1} and~\ref{tab:net_params2}, respectively.
\textcolor{black}{With the above parameter settings, the LSTM-based predictor requires approximately $5.9 M$} trainable parameters, whereas the TCN-based predictor contains only approximately $1.2M$ parameters, corresponding to about $20.2\%$ of the LSTM model size. 

\begin{table}[t]
	\caption{Parameter Configurations of the LSTM Predictor}
	\label{tab:net_params1}
	\centering
	\begin{tabular}{|c|c|c|}
		\hline
		\textbf{Parameter} & \textbf{Symbol}& \textbf{Value} \\
		\hline
		Hidden state dimension
		& $R$ & 256\\
		\hline
		Stacked LSTM layer number& $K$ & 4\\
		\hline
		Input length & $I$ & $\{2,4\}$\\
		\hline
		Prediction length
		& O & $\{10,8\}$\\
		\hline
		Direction mode & -- & Bidirectional\\
		\hline
		Batch size
		& -- & 32\\
		\hline
	\end{tabular}
\end{table}
\begin{table}[t]
	\caption{Parameter Configurations of the TCN Predictor}
	\label{tab:net_params2}
	\centering
	\begin{tabular}{|c|c|c|}
		\hline
		\textbf{Parameter} & \textbf{Symbol} & \textbf{Value}\\
		\hline
		Feature dimension & $H_\text{ide}$ & 128\\
		\hline
		Multiple-TCRBs number & $\text{HL}$ & 7\\
		\hline
		Kernel size & $k$ & 5\\
		\hline
		Batch size & $-$ & 32\\
		\hline
		
	\end{tabular}
\end{table}

\begin{figure}
	\centering
	\includegraphics[width=0.45\textwidth]{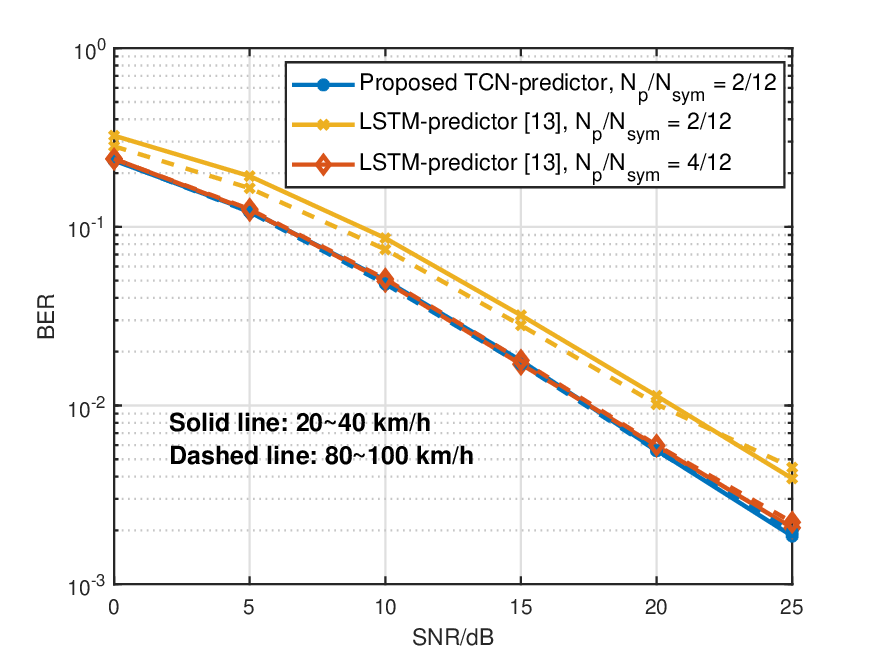}
	\caption{The BER performance comparison among different predictors under various Doppler conditions.}\label{fig:4}
\end{figure}

We first evaluate the BER performance of the two predictors under various Doppler conditions, as illustrated in Fig.~\ref{fig:4}. Specifically, both predictors are employed in the proposed transmission scheme shown in Fig.~\ref{fig:sys}, under medium-speed and high-speed mobility scenarios.
It can be observed that both predictors maintain nearly consistent performance across different SNR and Doppler conditions, demonstrating their ability to effectively capture the AR characteristics of CSI in the AF domain and their strong generalization capability for practical deployment.
Furthermore, the simulation results indicate that the TCN-based predictor significantly outperforms the LSTM-based one, achieving comparable BER performance with only two pilots to that of the LSTM with four pilots. This superiority is primarily attributed to the TCRB architecture. Specifically, the dilated convolutions in TCN provide an exponentially expanding receptive field, enabling more effective capturing of the AR characteristics of the channel compared to the recursive processing of LSTM. Furthermore, the residual connections within the TCRB establish identity mapping paths, allowing the network to focus on learning temporal variations rather than the entire channel state, which significantly enhances prediction accuracy under sparse pilot patterns. 

%\begin{figure*}[t]
%	\centering
%	\subfigure[BER performance comparison at 30-60 km/h.]{
%		\includegraphics[width=0.4\textwidth]{fig_nn_ber_cmp_3060}
%	}
%	\subfigure[BER performance comparison at 135-165 km/h.]{
%		\includegraphics[width=0.4\textwidth]{fig_nn_ber_cmp_135165}
%	}
%	\caption{BER performance comparison among different predictors under various Doppler conditions.}\label{fig:4}
	
%\end{figure*}

\begin{figure}
	\centering
	\includegraphics[width=0.45\textwidth]{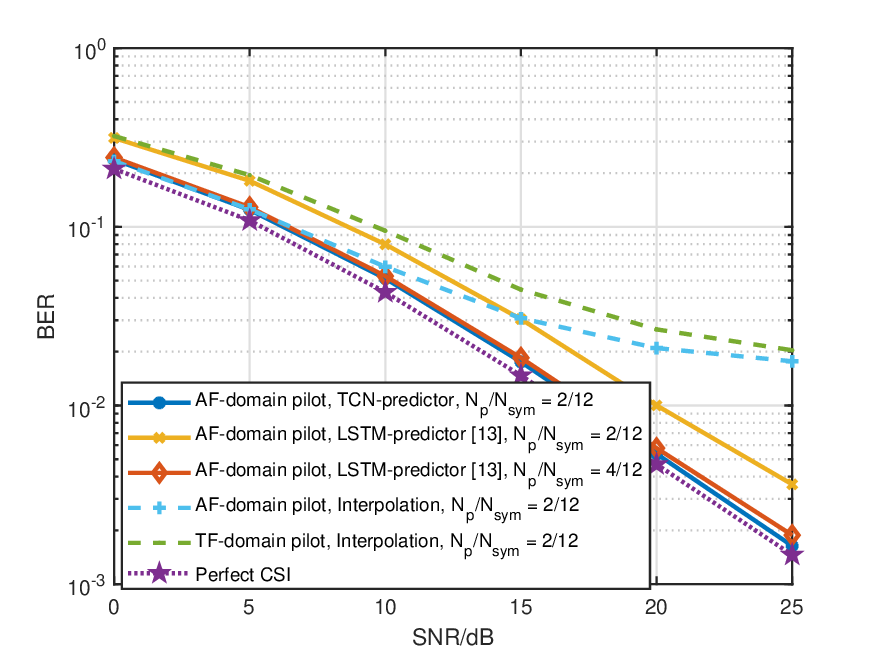}
	\caption{The BER performance comparison among different OFDM schemes.}\label{fig:5}
\end{figure}

Moving a step further, Fig. \ref{fig:5} compares the BER performance of various schemes across a velocity range of 0 to \textcolor{black}{100} km/h, highlighting the robustness and generalizability of the proposed framework. It is observed that the proposed schemes significantly outperform the conventional TF-domain pilot-based OFDM method under identical SNR conditions. This performance gain is attributed to the AF-domain channel estimation based on (\ref{AO-relation}), which effectively captures the ICI components that are often overlooked by traditional frequency-domain estimators. However, in the high-SNR regime, schemes without channel prediction (both TF and AF-domain) exhibit a noticeable error floor, primarily stemming from interpolation inaccuracies under high Doppler spreads. In contrast, the predictor-enhanced scheme effectively mitigates this error floor by leveraging the AR properties of the received pilots to achieve superior tracking accuracy. Notably, the predictor-enabled scheme closely approximates the perfect CSI baseline, demonstrating its significant potential for high-mobility communications.

We finally analyze the computational complexity of the LSTM-based and TCN-based predictors, respectively. Notably, the analysis focuses on the forward propagation complexity, i.e., the application complexity, since the network training can be performed offline.
%The forward propagation computational overhead of the proposed multi-layer LSTM predictor is mainly dominated by the recurrent gated operations and subsequent linear transformations. For a single LSTM layer with input $\mathbf{a}_{k}\in\mathbb{R}^{2N\times1}$ and hidden state $\mathbf{h}_{k}\in\mathbb{R}^{R\times1}$, the per-time-step complexity is $\mathcal{O}(4R(2N+R))$. Considering a sequence length $I$, $K$ stacked layers, and a bidirectional structure, the overall complexity of the recurrent module becomes $\mathcal{O}(2KIR(2N+R))$. The FCN for feature transformation and prediction introduces an additional cost of $\mathcal{O}(R^2+2RNO)$, which is negligible compared to the recurrent part.
%Therefore, the total complexity can be summarized as $\mathcal{O}((2KI+1)R^2 + (2ON+4KIN)R)$. 
%For the TCN-based predictor, the computational complexity is mainly determined by the stacked dilated convolutional layers and point-wise projections. After reshaping, the feature matrix is represented as $\mathbb{R}^{T_{\text{sym}}\times H_{\text{idden}}}$. For a 1D convolution with kernel size $k$, channel dimension $H_{\text{idden}}$, and sequence length $T_{\text{sym}}$, the complexity is $\mathcal{O}(T_{\text{sym}} k H_{\text{idden}}^2)$. Since each TCRB contains two such layers, a network with $\text{HL}$ blocks has a total complexity of $\mathcal{O}(2\cdot\text{HL}\cdot kT_{\text{sym}} H_{\text{idden}}^2 + 5NT_{\text{sym}}H_{\text{idden}})$.
The forward propagation computational overhead of the proposed multi-layer LSTM predictor is mainly dominated by the recurrent gated operations and subsequent linear transformations. For a single LSTM layer with input $\mathbf{a}_{k}\in\mathbb{R}^{2N\times1}$ and hidden state $\mathbf{h}_{k}\in\mathbb{R}^{R\times1}$, the per-time-step complexity is $\mathcal{O}(4R(2N+R))$. Considering a sequence length $I$, $K$ stacked layers, and a bidirectional structure, the overall complexity of the recurrent module becomes $\mathcal{O}(8KIR(2N+R))$. The FCN for feature transformation and prediction introduces an additional cost of $\mathcal{O}(2R^2+2RNO)$, which is negligible compared to the recurrent part.
Therefore, the total complexity can be summarized as~\cite{tang2026dopplerresilientleosatelliteofdm}
\begin{equation} \label{O1}
	\mathcal{O}\left((8KI+2)R^2 + (16KIN+2ON)R\right).
\end{equation}
For the TCN-based predictor, the computational complexity is mainly determined by the stacked dilated convolutional layers and point-wise projections. After reshaping, the feature matrix is represented as $\mathbb{R}^{T_{\text{sym}}\times H_{\text{ide}}}$. For a 1D convolution with kernel size $k$, channel dimension $H_{\text{ide}}$, and sequence length $T_{\text{sym}}$, the complexity is $\mathcal{O}(T_{\text{sym}} k H_{\text{ide}}^2)$. Since each TCRB contains two such layers, a network with $\text{HL}$ blocks has a total complexity of
\begin{equation} \label{O2}
	\mathcal{O}(2\cdot\text{HL}\cdot kT_{\text{sym}} H_{\text{ide}}^2 + 5NT_{\text{sym}}H_{\text{ide}}).
\end{equation}

By substituting the values from Table~\ref{tab:net_params1} and Table~\ref{tab:net_params2}, the inference complexity of the TCN-based predictor is found to be approximately twice that of the LSTM-based one. To further illustrate the computational burden of different predictors, we conduct multiple experiments to measure the inference time, and the detailed results are reported in Table~\ref{tab:net_runtime3}. It can be observed that the average inference time of the TCN-based predictor is indeed about twice that of the LSTM-based counterpart.
\textcolor{black}{These results indicate that, although the TCN-based predictor is significantly more lightweight in terms of model size, it achieves better performance in Fig.~\ref{fig:4} and Fig.~\ref{fig:5} at the expense of a higher computational cost during inference, highlighting the trade-off between inference latency and prediction accuracy. Nevertheless, considering the constrained radio resource, the TCN-based predictor is still preferred as it achieves the same BER performance with half pilot overhead.}
%These results indicate that, although the \textcolor{blue}{much more lightweight} TCN-based predictor achieves better performance in Fig.~\ref{fig:4} and Fig.~\ref{fig:5}, it incurs a higher computational cost, highlighting the trade-off between inference latency and prediction accuracy. 

\begin{table}[t]
	\caption{Inference time of Various Predictors}
	\label{tab:net_runtime3}
	\centering
	\begin{tabular}{|c|c|c|}
		\hline
		\textbf{Predictors} & \textbf{Average (ms)}& \textbf{Std (ms)} \\
		\hline
		TCN
		& 1.2148 & 0.0661\\
		\hline
		LSTM, $I=2$& 0.5832 & 0.0392\\
		\hline
		LSTM, $I=4$& 0.5652 & 0.0718\\
		\hline
	\end{tabular}
\end{table}
%In practice, $R$ is typically set as $R=\alpha_0 N$ with $\alpha_0\in\mathbb{Z}^+$, leading to an overall complexity dominated by $\mathcal{O}(N^2)$. Moreover, due to the sequential nature of LSTM, computations across time steps cannot be parallelized, resulting in increased inference latency.
% In practical implementations, $H_{\text{idden}}$ is typically set as $H_{\text{idden}}=\beta_0 N$ with $\beta_0\in\mathbb{Z}^+$, yielding an overall complexity that also scales as $\mathcal{O}(N^2)$. However, unlike LSTM, the convolutional operations can be fully parallelized across the time dimension, resulting in significantly improved computational efficiency and reduced inference latency.

\section{Conclusion}
\label{Conclusion}
In this paper, we proposed a novel AF-domain pilot-assisted OFDM framework for high-mobility LAWNs. By analyzing the relationship between frequency-domain and AF-domain channel representations, the proposed framework enables accurate frequency-domain CSI reconstruction and effectively characterizes Doppler-induced ICI on a path-wise basis.
To enhance channel estimation performance, we developed an RCG-based AF-domain pilot optimization scheme by minimizing the mutual coherence of the sensing dictionary. The optimized pilot effectively mitigates off-grid effects and improves sparse channel estimation performance without increasing the receiver complexity.
Furthermore, by exploiting the inherent AR structure of received pilot symbols, a TCN-based pilot prediction scheme was proposed to track rapidly time-varying channels and reduce estimation errors caused by high Doppler shifts.
Simulation results demonstrated that the proposed AF-domain pilot significantly outperforms conventional OFDM pilots in high-Doppler scenarios and achieves BER performance close to the perfect CSI benchmark, verifying the effectiveness and robustness of the proposed transmission framework.

\bibliographystyle{IEEEtran}%
\bibliography{bib/bibfile}

\vfill

\end{document}